\documentclass[a4paper,11pt]{article}

\pdfoutput=1 

\usepackage{jcappub} 

\usepackage{amsmath}
\usepackage{amssymb}
\usepackage{graphicx,caption,subcaption}

\makeatletter
\def\@hangfrom#1{\setbox\@tempboxa\hbox{{#1}}%
      \hangindent 0pt
      \noindent\box\@tempboxa}
\makeatother
\def\un#1{\relax\ifmmode\@@underline#1\else
        $\@@underline{\hbox{#1}}$\relax\fi}
\let\du=\du                     
\def\a{\alpha}
\def\b{\beta}

\def\d{\delta}

\def\g{\gamma}

\def\k{\kappa}
\def\l{\lambda}
\def\m{\mu}
\def\n{\nu}

\def\r{\rho}

\def\t{\tau}

\def\L{\Lambda}

\def\bo{{\raise-.3ex\hbox{\large$\Box$}}}               
\def\pa{\partial}                                       
\def\TH{{\raise.2ex\hbox{$\displaystyle \bigodot$}\mskip-4.7mu \llap H \;}}
\def\face{{\raise.2ex\hbox{$\displaystyle \bigodot$}\mskip-2.2mu \llap {$\ddot
        \smile$}}}                                      

\def\abs#1{\left| #1\right|}                    
\def\leftrightarrowfill{$\mathsurround=0pt \mathord\leftarrow \mkern-6mu
        \cleaders\hbox{$\mkern-2mu \mathord- \mkern-2mu$}\hfill
        \mkern-6mu \mathord\rightarrow$}
\def\dvec#1{\vbox{\ialign{##\crcr
        \leftrightarrowfill\crcr\noalign{\kern-1pt\nointerlineskip}
        $\hfil\displaystyle{#1}\hfil$\crcr}}}           
\def\frac#1#2{{\textstyle{#1\over\vphantom2\smash{\raise.20ex
        \hbox{$\scriptstyle{#2}$}}}}}                   
\def\sfrac#1#2{{\vphantom1\smash{\lower.5ex\hbox{\small$#1$}}\over
        \vphantom1\smash{\raise.4ex\hbox{\small$#2$}}}} 
\def\bfrac#1#2{{\vphantom1\smash{\lower.5ex\hbox{$#1$}}\over
        \vphantom1\smash{\raise.3ex\hbox{$#2$}}}}       
\def\afrac#1#2{{\vphantom1\smash{\lower.5ex\hbox{$#1$}}\over#2}}    
\def\[{\lfloor{\hskip 0.35pt}\!\!\!\lceil}
\def\]{\rfloor{\hskip 0.35pt}\!\!\!\rceil}

\def\du#1#2{_{#1}{}^{#2}}

\def\udu#1#2#3{^{#1}{}_{#2}{}^{#3}}

\def\ha{{\fracmm12}}

\def\un{\underline}
\def\fracmm#1#2{{{#1}\over{#2}}}

\def\low#1{{\raise -3pt\hbox{${\hskip 0.75pt}\!_{#1}$}}}

\newskip\humongous \humongous=0pt plus 1000pt minus 1000pt

\newif\ifdtup

\def\({\left(}
\def\){\right)}

\def\beq{\begin{equation}}
\def\eeq{\end{equation}}
\def\bea{\begin{eqnarray}}
\def\eea{\end{eqnarray}}
\newcommand{\be}{\begin{equation}}
\newcommand{\ee}{\end{equation}}
\newcommand{\nbe}{\begin{equation*}}
\newcommand{\nee}{\end{equation*}}

\newcommand{\lb}{\label}

\title{\boldmath On the superstring-inspired quantum correction to the Starobinsky model of inflation}

\author[a,b,c,1]{Sergei V.~Ketov,}\note{The corresponding author.}
\author[d]{Ekaterina O.~Pozdeeva,}
\author[d]{Sergey Yu.~Vernov}

\affiliation[a]{Department of Physics, Tokyo Metropolitan University\\
1-1 Minami-ohsawa, Hachioji-shi, Tokyo 192-0397, Japan}
\affiliation[b]{Research School of High-Energy Physics, Tomsk Polytechnic University\\
2a Lenin Avenue, Tomsk 634028, Russian Federation}
\affiliation[c]{Kavli Institute for the Physics and Mathematics of the Universe (WPI)
\\The University of Tokyo Institutes for Advanced Study, Kashiwa 277-8583, Japan}
\affiliation[d]{Skobeltsyn Institute of Nuclear Physics, Lomonosov Moscow State University,\\
Leninskiye Gory~1, Moscow 119991, Russian Federation}

\emailAdd{ketov@tmu.ac.jp}
\emailAdd{pozdeeva@www-hep.sinp.msu.ru}
\emailAdd{svernov@theory.sinp.msu.ru}


\arxivnumber{2211.01546}

\abstract{Superstring/M-theory is the theory of quantum gravity that can provide the UV-completion to viable inflation models. We modify the Starobinsky inflation model by adding the Bel-Robinson tensor $T^{\mu\nu\lambda\rho}$ squared term proposed as the leading quantum correction inspired by superstring theory. The $(R+\fracmm{1}{6m^2}R^2 -\fracmm{\beta}{8m^6}T^2)$ model under consideration has two parameters: the inflaton mass $m$ and the string-inspired positive parameter $\beta$. We derive the equations of motion in the Friedmann-Lemaitre-Robertson-Walker universe and investigate its solutions. We find the physical bounds on the value of the parameter $\beta$ by demanding the absence of ghosts and consistency of the derived inflationary observables with the measurements of the cosmic microwave background radiation.
}

\begin{document}
\maketitle
\flushbottom

\section{Introduction}

The Starobinsky model of inflation \cite{Starobinsky:1980te} is described by the modified gravity
action~\footnote{We do not follow historical developments.}
\be \lb{starm}
S_{\rm Star.}[g_{\m\n}] = \fracmm{M^2_{\rm Pl}}{2}  \int d^4x\sqrt{-g} \left(R+\fracmm{1}{6m^2} R^2\right)
\ee
with the reduced Planck mass $M_{\rm Pl}=1/\sqrt{8\pi G_N}$ and the inflaton (scalaron) mass $m$, in terms of metric
$g_{\m\n}$ having the Ricci scalar curvature $R$, with the spacetime signature $(-,+,+,+)$. This model has an attractor-type
solution describing a quasi-de Sitter expansion of the universe with the slow-roll inflation and a "graceful exit" in the
Friedmann-Lemaitre-Robertson-Walker (FLRW) universe. Being proposed the long time ago, the Starobinsky inflationary model is in perfect agreement with the
recent measurements of the cosmic microwave background (CMB) radiation \cite{Planck:2018jri,BICEP:2021xfz,Tristram:2021tvh}. The only free parameter $m$ in Eq.~(\ref{starm}) is fixed by CMB measurements (COBE normalization) as
\begin{equation}\lb{mass}
m=1.3 \left(\fracmm{55}{N}\right)10^{-5}M_{\rm Pl}= 3.2 \left(\fracmm{55}{N}\right)10^{13}\, \rm{GeV},
\end{equation}
where $N$ is the number of e-foldings describing the duration of inflation.

During slow roll inflation the Hubble function is essentially determined by the $R^2$ term in the action (\ref{starm}), the inflaton (called scalaron) is given by the physical (scalar) excitation of the higher-derivative gravity and can be interpreted as the Nambu-Goldstone boson associated with
spontaneous breaking of the scale invariance of the $R^2$-gravity action, see e.g., Refs.~\cite{Ketov:2010qz,Ketov:2012yz,Ketov:2019toi} for a review of these features.

However, the Starobinsky inflation is {\it unstable} against adding {\it generic} terms of the higher-order in the spacetime curvature to the action (\ref{starm}). In other words, the Starobinsky inflation is ultra-violet (UV) {\it sensitive} against quantum gravity corrections that have to be added because of nonrenormalizability of the quantum field theory (\ref{starm}), while the higher-order curvature terms are expected to be relevant in the high-curvature regime of inflation. Some simple candidates for the higher-order curvature terms were studied in the context of modified gravity and  inflation in Refs.~\cite{Barrow:1988xh,Gottlober:1989ww,Berkin:1990nu,Amendola:1993bg,Saidov:2010wx,Huang:2013hsb,Castellanos:2018dub,Ivanov:2021chn}, including the vicinity of the Starobinsky model, with the restrictions on the parameters of the modified gravity models being found. Some nonlocal generalizations of the Starobinsky inflation model were proposed in Refs.~\cite{Koshelev:2016xqb,Koshelev:2020foq,Koshelev:2022olc}. However, fixing the structure of the higher-order curvature terms is possible only in quantum gravity, while it is also of practical importance for inflation phenomenology because expected improvements of precision CMB measurements may require refinements of the Starobinsky model predictions.

Superstring/M-theory \cite{Green:2012pqa} is the mathematically consistent theory of quantum gravity that can provide the UV-completion to viable inflation models and offer insights into the gravitational effective field theory (EFT) in ten or eleven spacetime dimensions, respectively. As regards the gravitational EFT in four spacetime dimensions, its derivation requires compactification of extra (hidden) dimensions and moduli stabilization in addition. The leading corrections to the Einstein-Hilbert gravity action in 10 or 11 dimensions may be obtained in several ways: either (i) via a perturbative quantum field theory computation of the renormalization group beta-functions in the two-dimensional supersymmetric non-linear sigma-model (NLSM) describing a propagation of a test superstring in a curved space-time with the subsequent (Zamolodchikov-type) action whose equations of motion give the vanishing beta-functions, or (ii) via an expansion of the superstring graviton scattering amplitudes in powers of the string parameter $\alpha'$ with the subsequent construction of the gravitational EFT for them  \cite{Metsaev:1986yb}, or (iii) from the M-theory compactification or dimensional reduction \cite{Green:1997di,Tseytlin:2000sf}.

As regards closed (type-II) superstrings, the terms quadratic and cubic in the spacetime (full) curvature tensor are absent in the gravitational EFT, while the leading correction is given by the {\it quartic} terms, see e.g., the four-loop beta-function of the two-dimensional supersymmetric NLSM, computed by Grisaru, van de Ven  and Zanon in Ref.~\cite{Grisaru:1986px}, the corresponding gravitational  EFT found by Grisaru and Zanon in Ref.~\cite{Grisaru:1986vi}, and Refs.~\cite{Deriglazov:1991tm,Ketov:2000dy}  for more details and further generalizations. In particular, the absence of terms cubic in the curvatures comes from the non-existence of their locally supersymmetric extensions in four spacetime dimensions
\cite{Ketov:2000dy}.

The gravitational EFT obtained from {\it perturbative} superstring/M-theory is the subject to ambiguities related to field redefinitions of spacetime metric \cite{Ketov:2000dy}. In particular, neither the $R^2$ term nor the coefficient in front of it can be fixed that way.
Nevertheless, the $R^2$ term with a positive coefficient can be added to the gravitational EFT because it is the only ghost-free term in the quadratically generated gravity, while its coefficient can be fixed by phenomenological considerations. Of course, this proposal goes beyond the perturbative superstring theory because the $R^2$ term is supposed to be considered nonperturbatively (with the extra physical degree of freedom given by scalaron), whereas the known leading superstring correction found in  Ref.~\cite{Grisaru:1986vi} is perturbative by its derivation.

Moreover, details of compactification and moduli stabilization are highly non-trivial, while no viable solution is available for our Universe besides anthropological considerations. Nevertheless, spontaneous compactification and moduli stabilization are possible in principle by using fluxes in hidden dimensions \cite{Douglas:2006es}, while the stabilization of string dilaton and axion is only possible by nonperturbative (instanton-type) corrections with respect to the string coupling constant, leading to a non-vanishing scalar potential for the dilaton and axion fields \cite{Alexandrov:2016plh}.

An attempt to get the quartic curvature terms in the four-dimensional gravitational EFT by dimensional reduction of the
effective M-theory action \cite{Green:1997di,Tseytlin:2000sf} in 11 dimensions was made  in Ref.~\cite{Iihoshi:2007vv} leading to
their structure in the form of the Bel-Robinson (BR) tensor squared. However, those terms alone cannot describe inflation because of  very low duration (just a few e-foldings). Since any viable inflationary model in modified gravity has to have the $R^2$-term \cite{Ketov:2019toi}, the improved Starobinsky-Bel-Robinson (SBR) modified gravity action in four dimensions was proposed in Ref.~\cite{Ketov:2022lhx} by combining the $R^2$ and the BR tensor squared terms. Assuming that string dilaton and axion are stabilized, we can ignore their kinetic terms and keep only their mass terms, while the linear coupling of dilaton and axion to gravity in string theory is known via the Euler density and the Pontryagin density, respectively, in the leading approximation. This directly leads to the SBR action, see Sec.~2.

Our paper is organized as follows.  In Sec.~2 we formulate the SBR gravity action.  The equations of motion are derived in Sec.~3 where we also find the nonlinear ordinary differential equation (ODE) for the Hubble function in the FLRW background. The relevant ODE solutions are derived both analytically and numerically in Sec.~4. The inflationary observables are discussed in Sec.~5, where we obtain the observational bounds on the size of the quantum corrections. Section 6 is our Conclusion.  Appendices A, B, C and D are devoted to mathematical details. We do not provide an introduction to inflation and an analysis of  metric perturbations because they can be easily found in many publications. The list of our references is limited to directly relevant papers.

\section{Setup}

The Bel-Robinson (BR) tensor in four spacetime dimensions is defined by \cite{Bel:1959uwe,Robinson:1959ev,Deser:1999jw}
\begin{eqnarray} \label{belr}
T^{\m\n\l\r} & \equiv  & R^{\m\a\b\l}R\udu{\n}{\a\b}{\r} +{}^*R^{\m\a\b\l}~{}^*R\udu{\n}{\a\b}{\r} \nonumber \\
 & = & R^{\m\a\b\l}R\udu{\n}{\a\b}{\r} +R^{\m\a\b\r}R\udu{\n}{\a\b}{\l}-\ha g^{\m\n}R^{\a\b\g\l} R\du{\a\b\g}{\r}~,
\end{eqnarray}
by analogy with the energy-momentum tensor of the Maxwell theory of electromagnetism.

In curved four-dimensional  spacetime, we use the lower case Greek letters for vector indices. Our notation and some basic equations of Riemann geometry are given in Appendix~\ref{Notations}. We define dual tensors with the help of Levi-Civita tensors, e.g.,
\begin{equation}\lb{duals}
{}^*R_{\m\n\l\r}=\fracmm{1}{2}E_{\m\n\a\b}R^{\a\b}_{{\,\,\,}{\,\,\,}\l\r}~,\quad E_{\m\n\l\r}=\sqrt{-g}\,\epsilon_{\m\n\l\r}~~,
\end{equation}
where $\epsilon_{\m\n\l\r}$ is the constant Levi-Civita symbol from Minkowski (flat) spacetime.

The SBR gravity action is defined by \cite{Ketov:2022lhx}
\be \lb{sbr}
S_{\rm SBR}[g_{\m\n}]  = \fracmm{M^2_{\rm Pl}}{2}\int d^4x\sqrt{-g} \left[
R +\fracmm{1}{6m^2}R^2 - \fracmm{\b}{8m^6} T^{\m\n\l\r}T_{\m\n\l\r}\right]~,
\ee
where we have introduced the dimensionless coupling constant $\beta>0$.~\footnote{The constant $\beta$ used here is different
from the $\beta$ parameter of Ref.~\cite{Ketov:2022lhx} by its normalization and sign. In the context of superstrings and
M-theory~\cite{Ketov:2022lhx}, the constant $\b$ must be positive.}

The BR tensor squared can be rewritten in terms of the Euler and Pontryagin densities squared by using the identities~\cite{Deser:1999jw}
\be \lb{deserid}
T^{\m\n\l\r}T_{\m\n\l\r}=\fracmm{1}{4}\left(P_4^2-E^2_4\right) =\fracmm{1}{4}\left(P_4+E_4\right)\left(P_4-E_4\right)~,
\ee
where the Euler and Pontryagin (topological) densities have been introduced in $D=4$ dimensions as
\be \lb{topd}
E_4={}^*R_{\m\n\l\r}{}^*R^{\m\n\l\r} \quad {\rm and} \quad P_4={}^*R_{\m\n\l\r}R^{\m\n\l\r}~,
\ee
respectively. It is worth noticing that the Euler density coincides with the Gauss-Bonnet (GB) term, $E_4={\cal G}$, see Appendix~\ref{EGB}. Therefore, we can rewrite
the SBR action (\ref{sbr}) to the form
\be \lb{sbrm}
S_{\rm SBR}[g_{\m\n}]  =\fracmm{M^2_{\rm Pl}}{2}\int d^4x\sqrt{-g} \left[
R +\fracmm{1}{6m^2}R^2 + \fracmm{\b}{32m^6}\left({\cal G}^2-P_4^2\right) \right]~,
\ee
thus establishing a connection to the modified $f(R,{\cal G})$ gravity theories.~\footnote{To the best of our knowledge, the $P_4$-terms were
never considered in the modified gravity literature.} In particular, the positive sign of $\beta$ is consistent with the physical requirement in the
$F({\cal G})$ modified theories of gravity, demanding the second derivative of the $F$-function to be positive~\cite{DeFelice:2008wz}.

The SBR action (\ref{sbrm}) does not include terms with  the spacetime derivatives of the curvature tensors, which would lead to new physical degrees of freedom, {\it cf.} Ref.~\cite{Cuzinatto:2018vjt}.

The classical actions (\ref{sbr}) and (\ref{sbrm}) can also be rewritten to the following form:
\begin{equation}\label{Slfields}
S_{\rm SBR}[g_{\m\n},\phi,\chi,\xi]=\fracmm{M^2_{\rm Pl}}{2}S_{R}-\fracmm{M^2_{\rm Pl}\beta}{32m^6}\left(S_{\cal G}+S_{P}\right),
\end{equation}
where we have introduced the auxiliary scalar fields $\phi$, $ \chi$ and $\xi$, together with
\begin{eqnarray}
S_{R}[g_{\m\n},\phi]&=&\int d^4x\sqrt{-g}\left[R\left(1+\fracmm{\phi}{3 m^2}\right)-\fracmm{\phi^2}{6m^2}\right], \lb{Raction}\\
S_{\cal{G}}[g_{\m\n},\chi]&=&\int d^4x\sqrt{-g}\left(\fracmm{\chi^2}{2}-{\cal G}\chi\right),\lb{Gaction}\\
S_{P}[g_{\m\n},\xi]&=&\int d^4x\sqrt{-g}\left(\xi P_4-\fracmm{\xi^2}{2}\right). \lb{Paction}
\end{eqnarray}

Varying the action (\ref{Slfields}) with respect to the scalar fields, we get the equations
\begin{equation} \lb{seom}
\chi={\cal G},\quad \phi=R,\quad \xi=P_4~,
\end{equation}
while their substitution into Eq.~(\ref{Slfields}) yields back the action (\ref{sbrm}). The linearization of the SBR action (\ref{sbrm}) with respect to $R$, ${\cal G}$ and $P_4$ in the actions (\ref{Raction}), (\ref{Gaction}) and (\ref{Paction}), respectively,  allows us to identify $\phi$ with Starobinsky's scalaron, $\chi$ with string dilaton, and $\xi$ with string axion, in the presence of their mass terms and the apparent absence of their kinetic terms, {\it cf.} Ref.~\cite{DeFelice:2009ak}. Extra terms in the axion potential
are needed for its stabilization.

String theory is known to be defined only on Ricci-flat spacetime backgrounds. Therefore, the higher-order terms in the gravitational effective action with the full curvature tensors only are not ambiguous, whereas the terms including the Ricci curvatures are ambiguous under metric field redefinitions \cite{Ketov:2000dy}. The BR-term in the SBR gravity action (\ref{sbrm}) comes from string theory, and it is not ambiguous, {\it cf.} Ref.~\cite{Maroto:1997aw}.

\section{Equations of motion}

A variation of the action $S_{\rm SBR}$ in Eq.~(\ref{Slfields}) with respect to metric variations $\delta g_{\rho\nu}$ is given by
\begin{equation}\label{deltaS_l}
    \delta S_{\rm SBR}=\fracmm{M^2_{Pl}}{2}\delta S_R-\fracmm{M^2_{Pl}\beta}{32m^6}\left(\delta S_{\cal G}+\delta S_{P}\right),
\end{equation}
where we have (see Appendix~\ref{Equa}  for details)
\begin{equation}
\label{deltaSR}
\begin{split}
{\delta S_R}=&\!\int\!d^4x\sqrt{-g}\left[\left(R_{\rho\nu}-\fracmm{g_{\rho\nu}}{2}R\right)\left(1+\fracmm{\phi}{3m^2}\right)+\fracmm{\phi^2}{12m^2}g_{\rho\nu}\right.\\
-&\left.\fracmm{1}{3m^2}\left(\fracmm{\nabla_\rho\nabla_\nu+\nabla_\nu\nabla_\rho}{2}-g_{\rho\nu}\Box\right)\phi\right]\delta g^{\rho\nu},
\end{split}
\end{equation}
\be
\label{deltaSG}
\begin{split}
{\delta S_{\cal G}}=&\int d^4x\sqrt{-g}\left[{}-\fracmm{\chi^2}{4}g_{\rho\nu}-2\left\{[Rg_{\rho\nu}-2R_{\rho\nu}]\Box\chi-R\nabla_\rho\nabla_\nu\chi\right.\right.\\
    +&\left.\left.2(R^\alpha_\nu\nabla_\alpha\nabla_\rho\chi+R^\alpha_\rho\nabla_\alpha\nabla_\nu\chi)-2(g_{\rho\nu}R_{\alpha\beta}+R_{\alpha\rho\nu\beta})\nabla^\beta\nabla^\alpha\chi\right\}\right]\delta g^{\rho\nu}~.\\
\end{split}
\end{equation}

As regards a variation $\d S_P$, we find
\begin{eqnarray}  \lb{varP}
\int d^4x \sqrt{-g}\xi{\delta P_4} & = & \int d^4x \sqrt{-g}\left[ E_{\a\nu\rho\sigma} g_{\l\mu}\nabla^{\a}\nabla_{\k}(\xi R^{\rho\sigma \k\l})+E_{\a\mu\rho\sigma} g_{\l\nu}\nabla^{\a}\nabla_{\k}(\xi R^{\rho\sigma \k\l}) \right. \nonumber\\
 & - &{} \fracmm{\xi}{4}g_{\mu\nu}E_{\a\b\rho\sigma}R^{\rho\sigma}_{\quad \k\l}R^{\a\b\k\l}+\fracmm{\xi}{2}E_{\a\b\rho\sigma}
 (g^{\l\g} R^{\rho\sigma}_{\quad\mu \l}R^{\a\b}_{\quad\nu\g}+g^{\k\g} R^{\rho\sigma}_{\quad \k\nu}R^{\a\b}_{\quad\g\mu}) \nonumber\\
 & + & \left. \fracmm{\xi}{2}(E_{\a\nu\rho\sigma}R^{\rho\sigma \k\l}R^{\a}_{\,\,\mu \k\l}+E_{\a\mu\rho\sigma}R^{\rho\sigma \k\l}R^{\a}_{\,\,\nu \k\l})\right]{\delta g^{\mu\nu}}~.
\end{eqnarray}
In the rest of our paper, we are going to apply the SBR theory to inflation in a spatially flat Friedman-Lemaitre-Robertson-Walker (FLRW) universe characterized by the metric
\begin{equation}
\label{metric}
ds^2={}-dt^2+a^2\left(dx_1^2+dx_2^2+dx_3^2\right)
\end{equation}
with the cosmic scale factor $a(t)$. We find that the $P_4$ term in the SBR action does {\it not} contribute to the equations of motion in the FLRW case, so that we ignore contributions from the action $S_P$ in what follows. Then the equations of motion from the SBR action  are
\begin{equation}
\label{SYSTEM}
\begin{split}
  &\left(R_{\rho\nu}-\fracmm{g_{\rho\nu}}{2}R\right)\left(1+\fracmm{\phi}{3m^2}\right)+\fracmm{\phi^2}{12m^2}g_{\rho\nu}
-\fracmm{1}{3m^2}\left(\fracmm{\nabla_\rho\nabla_\nu+\nabla_\nu\nabla_\rho}{2}-g_{\rho\nu}\Box\right)\phi\\
&{}+\fracmm{\beta}{64m^6}\left[\chi^2g_{\rho\nu}+8\left\{(Rg_{\rho\nu}-2R_{\rho\nu})\Box\chi-R\nabla_\rho\nabla_\nu\chi\right.\right. \\
    &{}+\left.\left.2(R^\alpha_\nu\nabla_\alpha\nabla_\rho\chi+R^\alpha_\rho\nabla_\alpha\nabla_\nu\chi)-2(g_{\rho\nu}R_{\alpha\beta}+R_{\alpha\rho\nu\beta})\nabla^\beta\nabla^\alpha\chi\right\}\right]=0~,\\
    & \chi={\cal G}~,\\
    &\phi=R~.
\end{split}
\end{equation}
In particular, the trace equation reads
\begin{equation}\label{Equtr}
    -R\left(1+\fracmm{\phi}{3m^2}\right)+\fracmm{\phi^2}{3m^2}+\fracmm{1}{m^2}\Box\phi +\fracmm{\beta}{16m^6}\left[\chi^2+2(R\Box\chi-2R^{\alpha\beta}\nabla_\alpha\nabla_\beta\chi)\right]=0.
\end{equation}

The FLRW metric also leads to simplifications of the covariant derivatives as (no sums)
\begin{equation}\label{Christofell}
    \Gamma^{0}_{jj}=a\dot{a},\qquad \Gamma^{j}_{j0}=\Gamma^{j}_{0j}=\fracmm{\dot{a}}{a}\equiv H,
\end{equation}
\begin{equation}
\label{GB}
 {\cal G}=\fracmm{24(\dot{a})^2\ddot{a}}{a^3}=24H^2\left(\dot{H}+H^2\right) \quad\mbox{\rm and}\quad \sqrt{-g}\,{\cal G}=
 8\fracmm{d}{dt}\left({\dot{a}}^3\right)~,
\end{equation}
\begin{equation}\label{R}
    R=\fracmm{6}{a^2}\left({\dot{a}}^2+a\ddot{a}\right)=6\left(\dot{H}+2H^2\right),\quad R_{00}={}-3\fracmm{\ddot{a}}{a}=3H^2-\fracmm{R}{2},\quad R_{jj}=2{\dot{a}}^2+a\ddot{a}~,
\end{equation}
\begin{equation}\label{Rmunualphabeta}
    R^0_{\,\,j0j}={}-R^0_{\,\,jj0}=a\ddot{a}~,\quad  R^j_{\,\,00j}=\fracmm{\ddot{a}}{a},\quad R^i_{\,\,jkl}={\dot{a}}^2\left(\delta^i_k\delta^j_l-\delta^i_l\delta^j_k\right)\,,
\end{equation}
where we have listed only the non-vanishing components of the Christoffel symbols and the curvature tensors, in the notation
$i,j,k,l=1,2,3$, with the Hubble function $H=\dot{a}/a$, and the dots denoting the time derivatives.

In particular, the second covariant derivatives of a scalar field $\Phi(t)$ are given by
\begin{equation}
\label{DD}
\nabla_0\nabla_0 \Phi = \ddot{\Phi},\quad \nabla_i\nabla_j \Phi={}-a\dot{a}\dot{\Phi}\delta_{ij},\quad \nabla_j\nabla_0 \Phi=\nabla_0\nabla_j \Phi={}-\fracmm{\dot{a}}{a}\dot{\Phi},\quad
\Box\Phi={}-\ddot{\Phi}-3H\dot{\Phi}.
\end{equation}

The $(0,0)$-component of the equations of motion (\ref{SYSTEM}) in the FLRW universe reads
\begin{equation}\label{EQU00}
    3H^2\left(1+\fracmm{R}{3m^2}\right)-\fracmm{R^2}{12m^2}+\fracmm{H\dot{R}}{m^2}
    =\fracmm{\beta}{64m^6}\left[{\cal G}^2-48H^3\dot{{\cal G}}\right]~.
\end{equation}

The trace equation (\ref{Equtr}) can be rewritten to the form
\begin{equation}\label{EtrF}
    R-\fracmm{1}{m^2}\Box R=R +\fracmm{1}{m^2}\left( \ddot{R} + 3H \dot{R}\right) =
    \fracmm{\beta}{16m^6}\left[{\cal G}^2-12\left(H^2\ddot{\cal G}+2H\dot{H}\dot{\cal G}+3H^3\dot{\cal G}\right)\right].
    \end{equation}
The Klein-Gordon terms on the left-hand-side of this equation demonstrate that the scalaron $\phi=R$ is a dynamical field of the mass $m$, as it should be.

Using the relations
\be \dot{R}=24\,{\dot{H}}\,H+6\,{\ddot{H}} \quad {\rm and} \quad
\dot{\cal G}= 48\,H \left( {H}^{2}+{\dot{H}} \right) {\dot{H}}+24\,{H}^{2} \left( 2\,H
{\dot{H}}+{\ddot{H}} \right)~~,
\ee
we rewrite Eq.~(\ref{EQU00})  in terms of the Hubble function $H(t)$ and its time derivatives  as
\be  \label{odeH}
2\left({m}^{4} +3\beta\,{H}^{4}\right) H{\ddot{H}}-\left(m^4-9\beta\,{H}^{4}\right) {\dot{H}}^{2}+6 \left( m^4
+3\beta\,{H}^{4}\right){H}^{2}{\dot{H}}-3\beta\,{H}^{8}+{m}^{6}{H}^{2}=0~.
\ee
This non-linear ordinary differential equation (ODE) is one of the main new results of this paper. It is consistent
with the results of Ref.~\cite{Starobinsky:1980te} in the case of $\beta=0$.

In terms of the e-foldings number $N=-\ln(a/a_{end})$ running backwards with time, the relations
\be
dN=-Hdt~,\quad
\dot{H}={}-H\fracmm{dH}{dN}\equiv {}-HH^{\prime} \quad {\rm and}\quad \ddot{H}=H{H^{\prime}}^2+H^2H^{\prime\prime}\,,
\ee
where the primes denote the derivatives with respect to $N$, allow us to transform Eq.~(\ref{odeH}) to the following non-linear ODE:
\be
\label{00N}
 2\left({m}^{4}+3\,\beta\,{H}^{4} \right) H {H}^{\prime\prime}+ \left({m}^{4} +
 15\beta\,{H}^{4} \right) {{H}^{\prime}}^{2}-6\,H \left( m^4-3\,\beta\,{H}^{4} \right) {H}^{\prime}-3\beta\,{H}^{6}+{m}^{6}=0.
\ee

Next, by using the relation
\begin{equation}
R=6\left(2H^2-HH'\right),
\end{equation}
we rewrite Eq.~(\ref{00N}) as the following dynamical system:
\begin{equation}\label{SySRHN}
\begin{split}
    H'&=2H-\fracmm{R}{6H}~~,\\
    R'&=\fracmm{1620\beta H^8-252\beta H^6R+9\beta H^4R^2+12m^4(3m^2+R)H^2-m^4R^2}{12H^2\left(m^4+3\beta H^4\right)}~~.
\end{split}
\end{equation}

The parameter $m$ can be removed from Eq.~\eqref{odeH} by using the dimensionless variables $h=H/m$ and $\tau=mt$, which yields
\begin{equation}
\label{equh}
 2h\left(1+3\beta h^{4}\right){\fracmm {d^{2}h}{d{\tau}^{2}}}- \left(1- 9\beta h^{4}\right) \left( {\fracmm {d h}{d\tau}} \right)^{2}+6h^{2} \left(1+3\beta h^{4}\right) {\fracmm {dh}{d\tau}} +h^{2} \left(1- 3\beta h^{6} \right)=0~.
\end{equation}

In turn, Eq.~(\ref{equh}) can be rewritten as
\begin{eqnarray}
\label{EquRt}
    \fracmm{dR_h}{d\tau}&=&\fracmm{1}{h(1+3\beta\, h^4)}\left[
 \fracmm{1}{12}R_h^2-R_h h^2-3h^2   -
    \fracmm{3}{4}\beta \left(R_h-10h^2\right)\left(R_h-18h^2\right)h^4\right]~,\\
 \label{Equht}
   \fracmm{dh}{d\tau}&=&\fracmm{R_h}{6}-2h^2,
\end{eqnarray}
after rescaling $R_h=R/m^2$. The System \eqref{SySRHN} is equivalent to
\begin{equation}\label{SySRh_h_N}
\begin{split}
    h'&=2h-\fracmm{R_h}{6h}~~,\\
    R_h'&=\fracmm{1620\beta h^8-252\beta h^6R_h+9\beta h^4R_h^2+12(3+R_h)h^2-R_h^2}{12h^2\left(1+3\beta h^4\right)}~~.
\end{split}
\end{equation}

Some solutions to these equations are derived in Sections 4 and 5.

\section{Solutions for FLRW background}

An exact  inflationary solution to Eq.~(\ref{odeH}) for the Hubble function in the FLRW universe is unknown in a finite form even in the case of $\beta=0$ but analytical approximations of the solution in the form of Laurent series are important for cosmological applications. Numerical solutions can also be used for that purpose, while the value of the parameter $\beta$ becomes important. We derive the physical conditions on the parameter $\beta$ in Sec.~5. Some formal mathematical properties (Painlev\'e tests) of Eq.~(\ref{odeH}) are obtained in Appendix~\ref{Painleve}.

\subsection{Solutions with $\b=0$}

First, we recall the well-known Starobinsky model of inflation \cite{Starobinsky:1980te}, arising from our model (\ref{sbrm})
in the case $\beta=0$. Our equation (\ref{odeH}) for the Hubble function in this case is consistent with the results of Ref.~\cite{Starobinsky:1980te}. Different aspects of the Starobinsky model of
inflation are described by Refs.~\cite{Ketov:2019toi,Ivanov:2021chn,Koshelev:2016xqb,Mishra:2019ymr,Ketov:2012jt}.

The $00$-component of the gravitational equations of motion (\ref{EQU00}) with $\b=0$ reads
\be \lb{scase}
12 H^2\left(R+3m^2\right) - R^2 + 12H\dot{R}=0~,
\ee
and the trace equation (\ref{EtrF}) with $\b=0$ gives the Klein-Gordon equation
\be \lb{kge}
\ddot{R} + 3H\dot{R} +m^2 R=0~.
\ee

Using the relation \eqref{R}, Eq.~(\ref{scase}) can be rewritten in terms of the Hubble function $H(t)$
 as the non-linear ODE of the 2nd order (dubbed the {\it Starobinsky equation})
\be  \lb{eomH}
2H\ddot{H} - \left(\dot{H}\right)^2 + H^2\left(6\dot{H} + m^2\right)=0~,
\ee
whereas Eq.~(\ref{kge}) appears to be of the 3rd order in terms of $H$ and its time derivatives.
It is easy to verify that Eq.~(\ref{kge}) is a consequence of Eq.~(\ref{eomH}).

In the slow-roll approximation defined by the conditions
\be  \lb{slowroll}
\abs{\ddot{H}} \ll \abs{H\dot{H}} \quad {\rm and} \quad \abs{\dot{H}} \ll H^2~,
\ee
Eq.~(\ref{eomH}) is greatly simplified to
\be \lb{sreomH}
6 \dot{H} + m^2\approx 0~,
\ee
and has the well-known solution
\be \lb{srsol}
H(t) \approx \fracmm{m^2}{6}(t_0-t)~,
\ee
where $t_0$ is the integration constant that apparently corresponds to the end of inflation, so that this leading term in
$H(t)>0$ should be a good approximation for $t\ll t_0$. The slow roll conditions (\ref{slowroll}) are valid provided that
$m(t_0-t)\gg 1$.

The slow-roll approximate solution (\ref{srsol}) for generic time values appears to be part of two different series of solutions.
On the one side, searching for a more general solution to Eq.~(\ref{eomH}) is possible by assuming its expansion in the form of {\it right\/} Painlev\'e series~\cite{Miritzis:2000js,Paliathanasis:2016tch}, with a movable singularity and the most singular term proportional to $c_p/(t-t_0)^p$ as
\begin{equation}\label{HLoran}
    H(t)=\sum\limits_{k=-p}^{+\infty}c_k(t-t_0)^k~~,
\end{equation}
where $c_k$ are constants, and $p$ is a natural number. Substituting Eq.~(\ref{HLoran}) into  Eq.~(\ref{eomH}), we get the following analytic  expansion:
 \begin{equation}
 \label{Pexpand}
 \begin{split}
H(t) & = {} -\fracmm{1}{2(t_0-t)}+\fracmm{m^2}{6}(t_0-t) - \fracmm{2m^4}{225}(t_0-t)^{3}+\fracmm{4}{4725}m^6(t-t_0)^5
\\
& + \fracmm{46}{1535625}m^8(t-t_0)^7+{\cal O}\left((t-t_0)^9\right)~.
\end{split}
 \end{equation}
The Ricci scalar curvature corresponding to the solution (\ref{Pexpand}) reads
\be \lb{curvL}
R=6(\dot{H}+2H^2)= \fracmm{3}{5} m^4 (t_0-t)^2  -3m^2 - \fracmm{4}{4875} m^8 (t_0-t)^6 +\ldots
\ee
and has no singular terms. The solution (\ref{Pexpand}) is the particular case of a more general  (non-attractor) solution in the form of Puiseux series with two integration constants, see Appendix~\ref{Painleve}.

The solution (\ref{Pexpand}) may be interpreted as a solution for inflation starting at a finite time and going for a short
time $t>t_0$ with the first term as the leading contribution, though with a singularity at $t=t_0$ and without a slow-roll regime. Being valid only for $m\abs{t-t_0}\ll 1$, Eq.~(\ref{Pexpand}) cannot be considered as an improvement of the slow-roll solution~(\ref{srsol}).

In the $(R+\fracmm{1}{6m^2}R^2 -2\L)$ modified gravity model with a cosmological constant $\L$, the equation of motion
(\ref{eomH}) is changed to
\be  \lb{eomHL}
2H\ddot{H} - \left(\dot{H}\right)^2 + H^2\left(6\dot{H} + m^2\right)=\fracmm{\L}{3} m^2~,
\ee
while it has the exact solutions given by~\cite{Vernov:2019ubo,Dimitrijevic:2021aio}
\be  \lb{eomHL1}
H_1 = {} - \fracmm{m^2}{6} (t-t_0) \quad {\rm with} \quad \L = {}-\fracmm{m^2}{12}
\ee
and
\be  \lb{eomHL2}
H_2 = \fracmm{1}{2(t-t_0)} - \fracmm{m^2}{6} (t-t_0) \quad {\rm with} \quad \L = {} -\fracmm{m^2}{3}~.
\ee
It implies that substituting the slow-roll solution (\ref{srsol}) into the left-hand-side of Eq.~(\ref{eomH}) yields a large non-vanishing constant instead of zero, while it is also true for substituting a sum of the singular and linear terms of the expansion  (\ref{Pexpand}). Hence, the singular term in Eq.~(\ref{Pexpand}) cannot be treated as the subleading term to the
approximate slow-roll solution~(\ref{srsol}).

In order to get an attractor (special) solution to Eq.~(\ref{odeH}), one should use the {\it ansatz} in the form of {\it left\/} Painlev\'e series~\cite{Miritzis:2000js,Paliathanasis:2016tch}
\begin{equation}
\label{HAntiLoran}
    H(t)=\sum^{k=p}_{k=-\infty}c_k(t_0-t)^k~.
\end{equation}
Then we find
\begin{equation}
\label{Hexpand}
\begin{split}
H(t)  & =  \fracmm{m^2}{6}(t_0-t)+\fracmm{1}{6(t_0-t)} - \fracmm{4}{9m^2(t_0-t)^3}+
\fracmm{146}{45m^4(t_0-t)^5}  \\
& {} -\fracmm{11752}{315 m^6 (t_0-t)^7} + {\cal O} \left((t_0-t)^{-9}\right)
\end{split}
\end{equation}
that is the true extension of the slow-roll solution (\ref{srsol}) by the subleading terms, being valid for $m(t_0-t)\gg 1$.
The first few terms of this expansion were found in Refs.~\cite{Starobinsky:1980te,Koshelev:2020foq} by using approximations to Eq.~(\ref{odeH}). Being a solution to the exact equation (\ref{odeH}), our expansion (\ref{Hexpand}) is in agreement with Eq.~(2.5) of Ref.~\cite{Koshelev:2020foq}. This solution is an attractor, it does not have a movable singularity
and describes (formally) eternal inflation up to $t\to -\infty$.

The Ricci scalar curvature corresponding to the solution (\ref{Hexpand}) reads
\begin{equation}
\lb{scurvS}
R(t)  = \fracmm{m^4}{3} (t_0-t)^2-\fracmm{m^2}{3} -\fracmm{4}{9(t_0-t)^2}+\fracmm{16}{5m^2(t_0-t)^4} - \fracmm{6908}{189m^4(t_0-t)^6}+ {\cal O} \left((t_0-t)^{-8}\right)~.
\end{equation}

To study viable inflationary scenarios, it is useful to change the independent variable in Eq.~(\ref{eomH}) from time $t$ to the e-folding number $N$, which yields
\be
\label{00Nbeta0}
 2H{H}^{\prime\prime}-6HH^\prime+{H^\prime}^2+m^2=0~.
\ee
Using the identity
\begin{equation*}
HH^{\prime\prime}=\left(H^\prime\right)^2+H^2\left(\fracmm{H^\prime}{H}\right)^\prime,
\end{equation*}
we rewrite it in the form
\be
2H^2\left(\fracmm{H^\prime}{H}\right)^\prime-6HH^\prime+3\left(H^\prime\right)^2+m^2=0~.
\ee

The slow-roll solution (\ref{srsol}) is given by
\be
\lb{srdiml}
H^2= \fracmm{m^2}{3}N~,
\ee
where $N$ is measured backwards from the end of inflation.~\footnote{Compared to Ref.~\cite{Ivanov:2021chn}, we include the integration constant $N_0$ into $N$.} It allows us to estimate the maximal value of the Hubble parameter $H$ during inflation by taking $N_{\rm max.}=65$ as
\be \lb{maxh}
H_{\rm max.} = 4.655m~.
\ee

\begin{figure}[htb]
 \centering
 \includegraphics[width=0.3\textwidth]{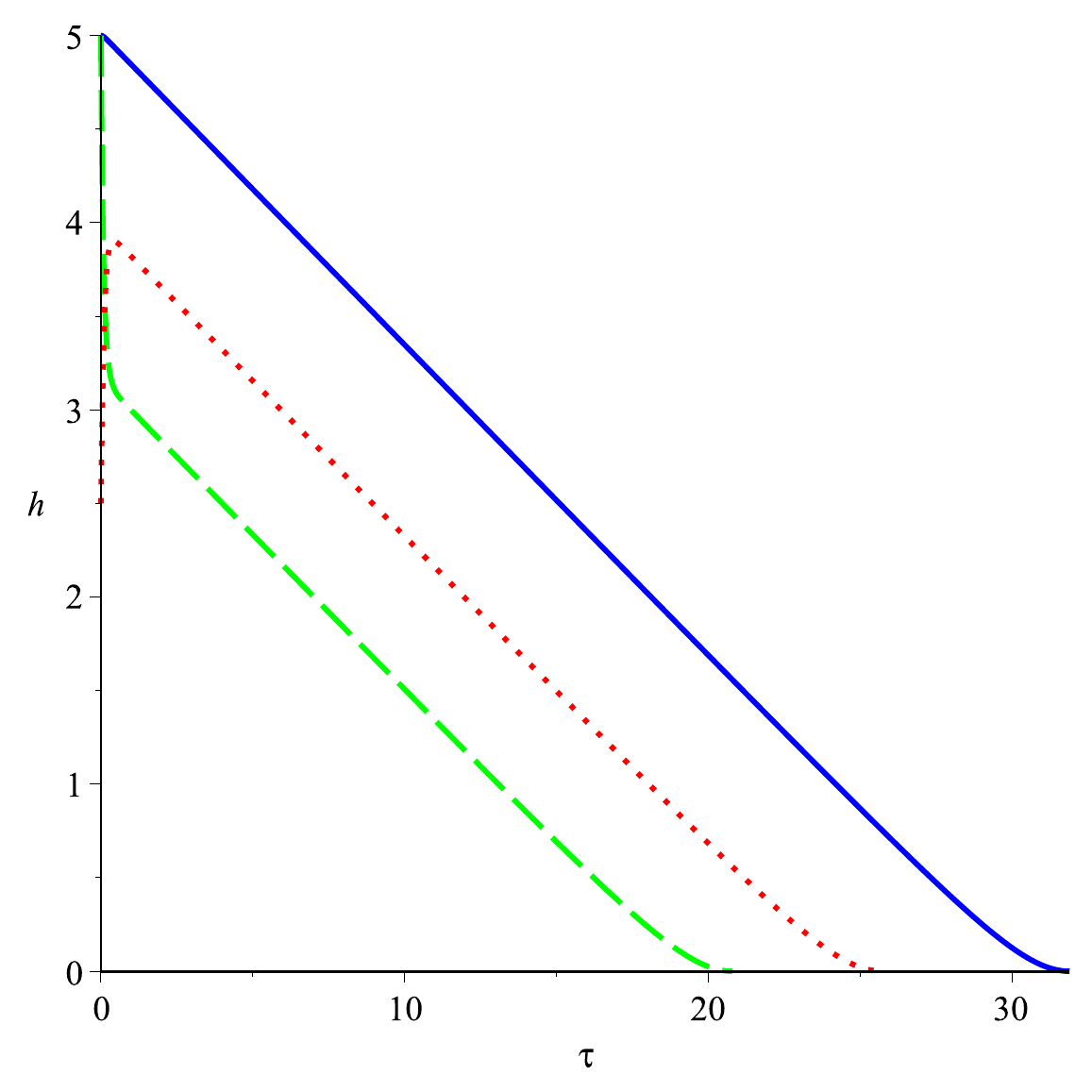} \
  \includegraphics[width=0.3\textwidth]{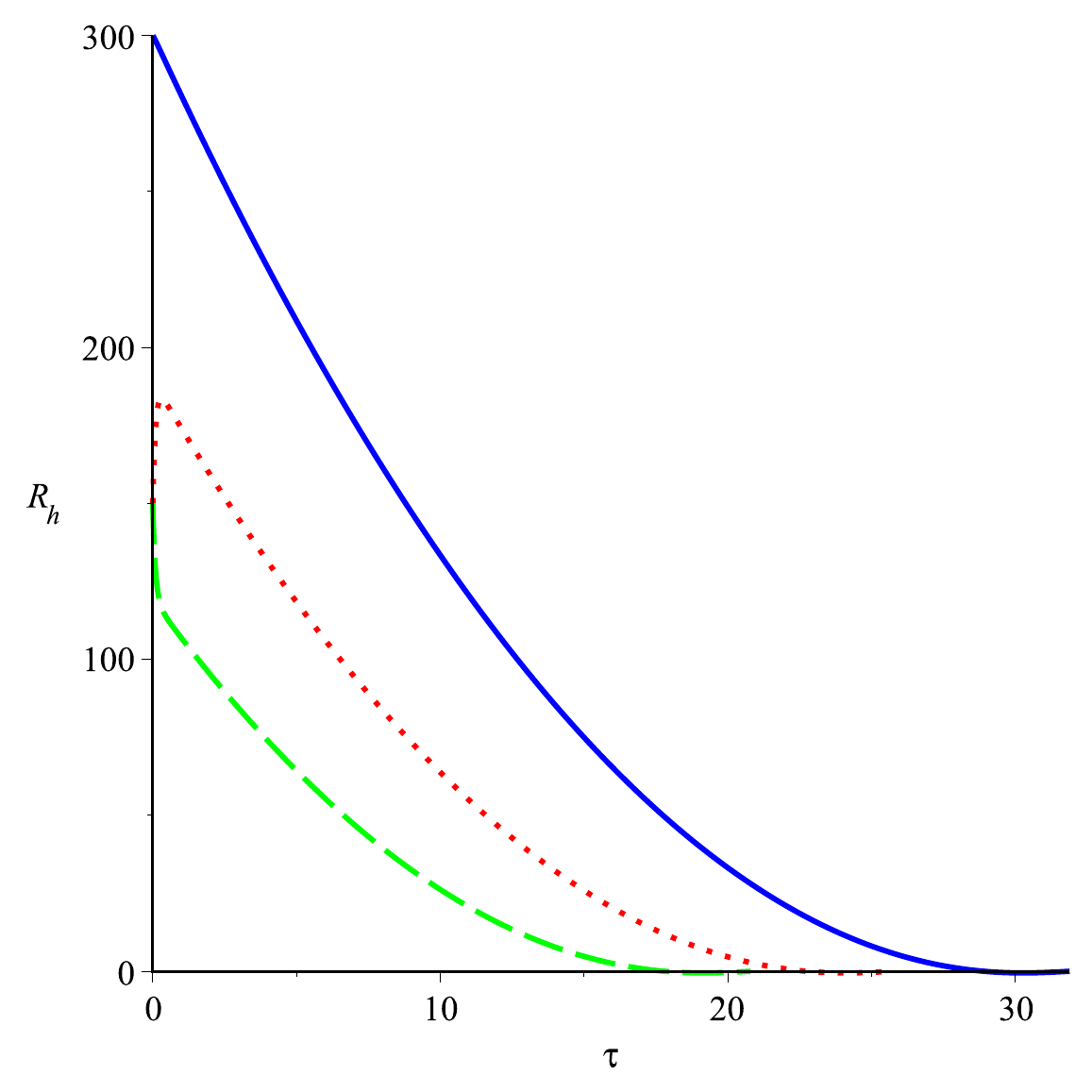} \
   \includegraphics[width=0.3\textwidth]{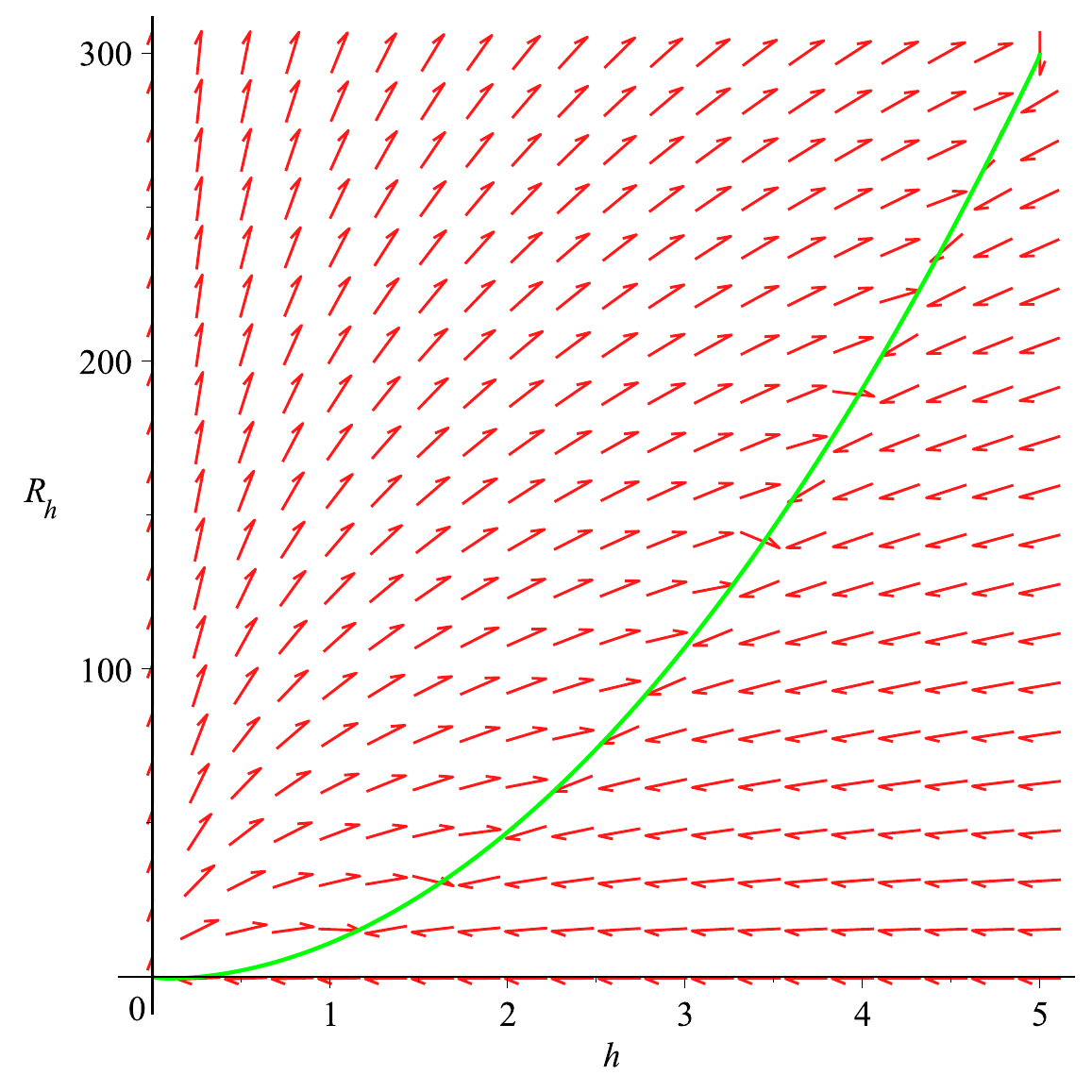}
\caption{The functions $h(\tau)$ (on the left), $R_h(\tau)$ (in the center) for different initial conditions, and the phase portrait $(R_h,h)$  (on the right) in the Starobinsky $(R+R^2)$ gravity model of inflation.}
\label{R2Solutionst}
\end{figure}

Numerical solutions to the exact dynamical systems (\ref{EquRt}) and (\ref{Equht}) for $\beta=0$ with some initial conditions are displayed in Fig.~\ref{R2Solutionst}. They are all the attractors, being close to those in the slow-roll approximate solution (\ref{srsol}).

\subsection{Solutions with a nonvanishing $\beta>0$}

The striking feature of Eq.~(\ref{odeH}) for $\beta>0$ versus the Starobinsky equation for $\beta=0$ is the existence of
a de Sitter exact solution with
\be \lb{deS}
H_{\rm dS} = mh_{\rm dS}=\fracmm{m}{(3\b)^{1/6}}={\rm const.}
\ee
that was already noticed in Ref.~\cite{Iihoshi:2007vv}. Substituting $h=h_{\rm dS} +\d h$ with a small $\t$-dependent perturbation $\d h$ into Eq.~(\ref{odeH}) leads to a linear equation on $\d h$ in the form
\be \lb{linpe}
\fracmm{d^2 (\d h)}{d\t^2} + \fracmm{3}{(3\b)^{1/6}}\fracmm{d(\d h)}{d\t} - \fracmm{3}{1+(3\b)^{1/3}}\d h =0~.
\ee
The general solution to Eq.~(\ref{linpe}) is given by
\be \lb{linsol}
\d h = c_{-} e^{\a_{-}\t} + c_{+} e^{\a_{+}\t}~,
\ee
where $c_{\pm}$ are the integration constants and $\a_{\pm}$  are the roots of a quadratic equation,
\be \lb{roots}
\a_{\pm} = -\fracmm{3}{2(3\b)^{1/6}} \left[ 1 \mp \sqrt{ 1+ \fracmm{4(3\b)^{1/3}}{3[1+(3\b)^{1/3}]}}~\right]~~.
\ee
In particular, when $\b^{1/3}\ll 1$, we have
\be\lb{sroots}
 \a_- \approx -\fracmm{3}{(3\b)^{1/6}} \quad {\rm and } \quad  \a_+\approx +(3\b)^{1/6}~~.
\ee
Since one of the roots is always positive the solution (\ref{deS}) is unstable.

To the end of this Subsection we derive quantum corrections to the Starobinsky inflation due to the $\beta$-dependent terms, assuming that the parameter $\beta$ is small, say, of the order $10^{-5}$ or less. The upper bounds on the parameter $\beta$ are obtained from various physical requirements in Sec.~5.

We search for a perturbative solution to Eq.~(\ref{odeH}) in the first order with respect to the parameter $\b$, in the
dimensionless form
\begin{equation} \lb{Hform}
h(\t)=\fracmm{1}{6}(\t_0-\t)-\beta h_1(\t),
\end{equation}
where we have kept only the leading term in the solution to the Hubble function of the Starobinsky inflation during the slow-roll regime, with the unknown function $h_1(\t)$.

In the first order with respect to $\beta$, we find a linear equation on $h_1$ in the form
\begin{equation}
\label{Eq00beta1}
   (\t_0-\t) \fracmm{d^2h_1}{d\tau^2}+\left[1+\fracmm{1}{2}(\t-\t_0)^2\right] \fracmm{dh_1}{d\tau}+\fracmm{1}{186624}(\t-\t_0)^4\left[(\t-\t_0)^4+36(\t-\t_0)^2-108\right]=0~.
\end{equation}
The general solution to Eq.~(\ref{Eq00beta1}) is given by
\begin{equation}
\label{h_1analitic}
\begin{split}
h_1 & =  {}-\fracmm{(\tau-\tau_0)^7}{653184}-\fracmm{23(\tau-\tau_0)^5}{233280}-\fracmm{7(\tau-\tau_0)^3}{11664} \\
& {}-\fracmm{7(\tau-\tau_0)}{1944}
+\fracmm{7\sqrt{\pi}}{1944}\rm{erf}\left(\fracmm{(\tau-\tau_0)}{2}\right)e^{(\tau-\tau_0)^2/4}+C_2+2C_1e^{(\tau-\tau_0)^2/4}~,
\end{split}
\end{equation}
where we have introduced the integration constants $C_1$ and $C_2$. The appearance of the exponential terms in this solution indicates instability of the Starobinsky inflation solution outside the slow-roll regime. We restrict ourselves to
the slow-roll approximation in what follows.

Taking into account the slow-roll conditions (\ref{slowroll})
allows us to simplify Eq.~(\ref{odeH}) to the non-linear ordinary differential equation
\be\label{eqbeta}
6 \left(m^4+3\beta H^{4}\right)\dot{H}-3\beta H^{6}+m^6=0~.
\ee
When searching for a perturbative solution to this equation in the first order with respect to $\b$, we find a simple answer,
\be\label{solbeta}
H(t)\approx \fracmm{m^2(t_0-t)}{6}- \beta\left( \fracmm{m}{6}\right)^6 (t_0-t)^{5} \left[ \fracmm{{m}^{2}}{14}(t_0-t)^{2}+\fracmm{18}{5} \right]~.
\ee
The first derivative of the Hubble function \eqref{solbeta} with respect to time reads
\be
\dot{H} ={} -\fracmm{{m}^{2}}{6}+{\fracmm {\beta\,{m}^{6}\,{(t-t_0)}^{4} \left[ {(t-t_0)}^{2}\,
{m}^{2}+36 \right]}{2^7\cdot 3^6}}~.
\ee

The slow-roll conditions \eqref{slowroll} also simplify the expressions for $\dot{\cal G}$  and ${\cal G}^2$ as follows:
\be\dot{\cal G}=96\, H^3\, \dot{H}+24\, H^2\, \ddot{H}+48\, H\, \dot{H}^2\approx96 H^3\dot{H},\ee
and
\be {\cal G}^2=576\,\left(H^8+2\,H^6\dot{H}+H^4\dot{H}^2\right)\approx576\,H^6\,\left(H^2+2\,\dot{H}\right)~. \ee
Accordingly, we find
\be
48 H^3\dot{\cal G}-{\cal G}^2\approx (24)^2 \left(6\dot{H}H^6-H^8\right)
\ee
that allows us to simplify Eq.~(\ref{EQU00}) as
\be\label{eqdH}
R\left(\fracmm{{R}}{12}-{H}^{2}\right)-H{\dot{R}}=3\,{m}^{2}\left(
{H}^{2}-{\fracmm {3\,\beta\,{H}^{8}}{{m}^{6}}}+{\fracmm {18\,\beta\,{H}^{6}{\dot{H}}}{{m}^{6}}}
\right)~~.
\ee

This equation can be further simplified by noticing that in the first order with respect to $\b$, we have
\be\label{prop}
\beta\dot{H}\approx\beta\dot{H}_0\approx-\fracmm{\beta\, m^2}{6}~.
\ee
Substituting Eq.~\eqref{prop} to Eq.~\eqref{eqdH}, yields
\be\label{Rs}
R\left(\fracmm{{R}}{12}-{H}^{2}\right)-H{\dot{R}}=3\,{m}^{2}\left({H}^{2}-{\fracmm {3\,\beta\,{H}^{8}}{{m}^{6}}}-{\fracmm {3\b
{H}^{6}}{{m}^{4}}}\right)~~.
\ee

\begin{figure}[htb]
 \centering
 \includegraphics[width=0.3\textwidth]{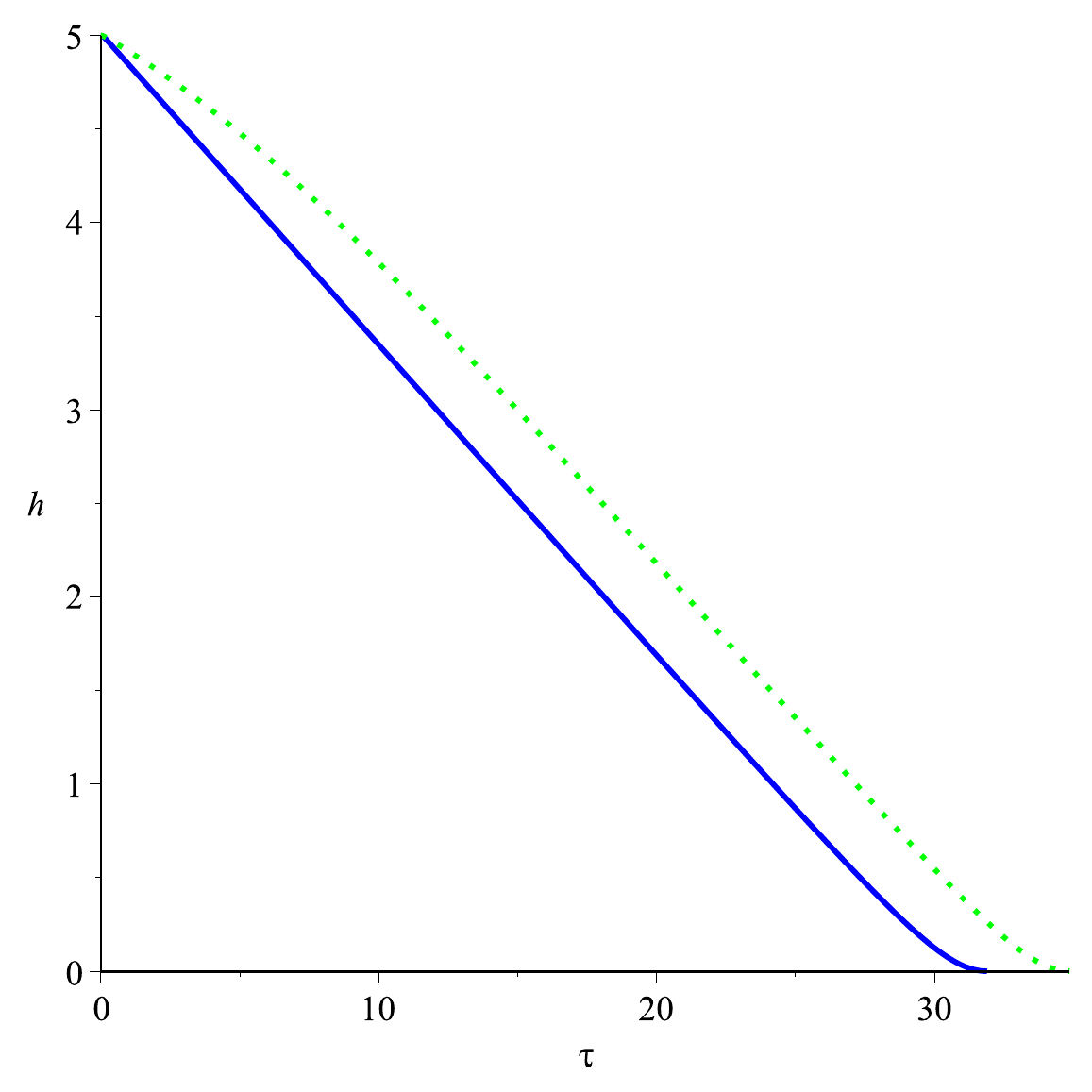} \
  \includegraphics[width=0.3\textwidth]{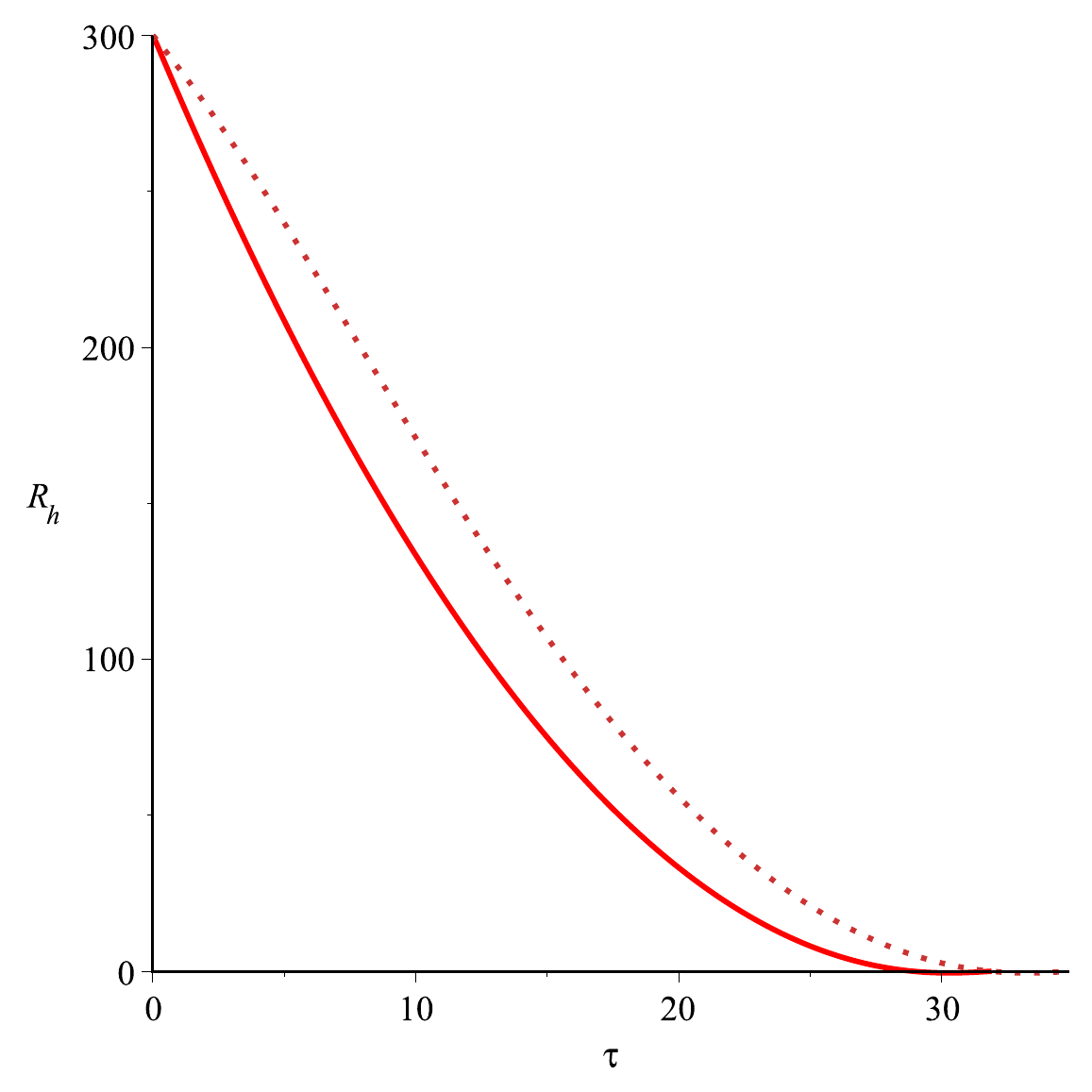} \
   \includegraphics[width=0.3\textwidth]{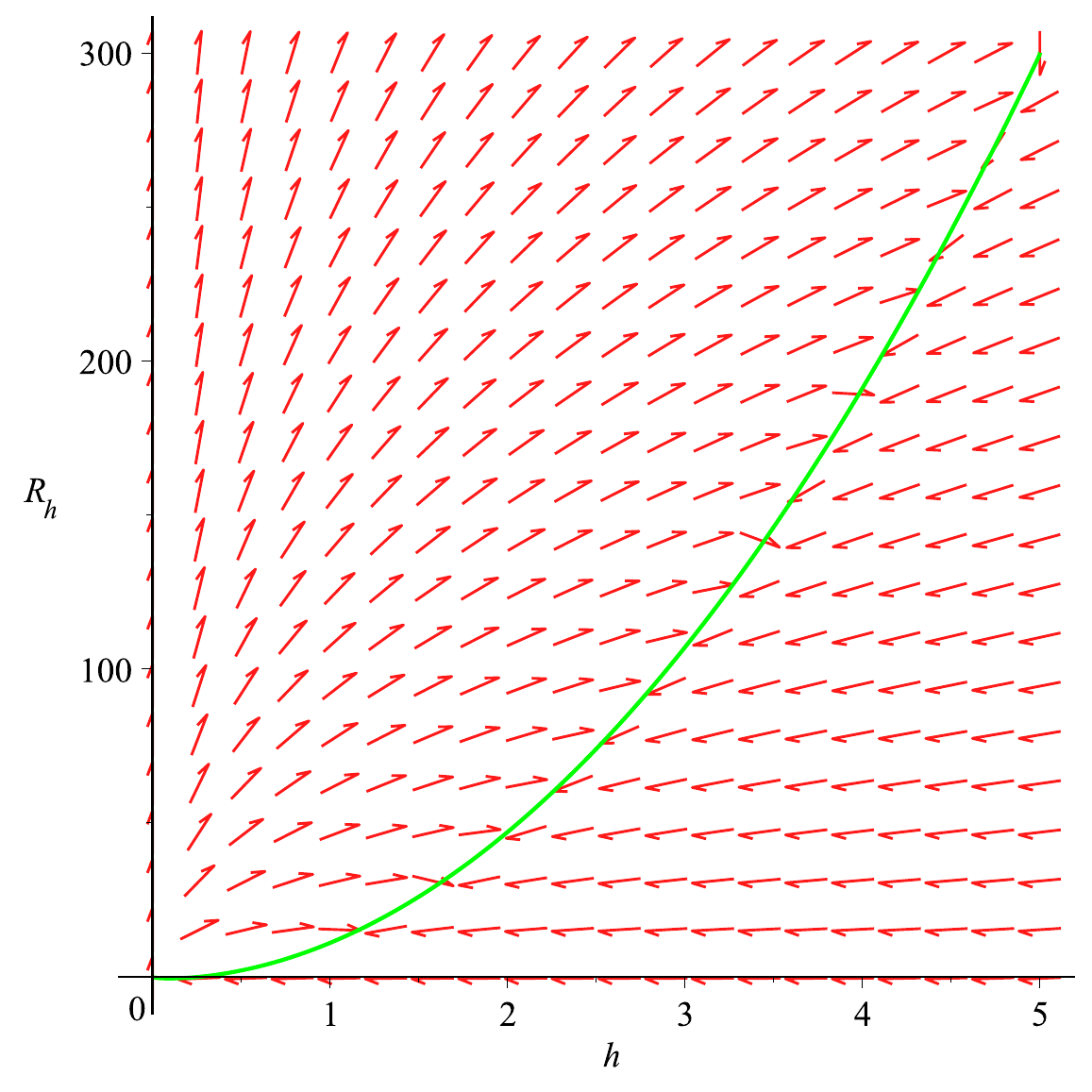}
\caption{The functions $h(\tau)$ (on the left), $R_h(\tau)$ (in the center) for $\beta=0$ (blue and red curves) and
$\beta=10^{-5}$ (green and brown  curves), and the phase portrait $(R_h,h)$ for $\beta=10^{-5}$ (on the right).}
\label{Solutionstbeta_m10m5}
\end{figure}

Numerical solutions to the dynamical systems (\ref{EquRt}) and (\ref{Equht}) for $\beta=10^{-5}$ are displayed
in Fig.~\ref{Solutionstbeta_m10m5}. They are close to those in the $\b=0$ case, see Fig.~\ref{R2Solutionst}.

In terms of e-folds $N$, the slow-roll solution to the Hubble function reads
\be\label{hbeta}
h = \sqrt{ \fracmm{N}{3}}+{\fracmm {\beta\,{N}^{5/2} \left( N+4 \right) }{72\sqrt {3}}} + {\cal O}\left(\b^2\right)~.
\ee

\section{Physical bounds on the value of $\beta$}

In Ref.~\cite{Cano:2020oaa} Cano, Fransen and Hertog studied various scenarios of inflation in the neighborhood of the Starobinsky model modified by the higher-order curvature terms, depending upon the unknown effective function $F(H^2)$ entering the equations of motion in the FLRW universe. Though our modification (\ref{sbrm}) of the Starobinsky model is outside their modified gravity theories because Eq.~(\ref{odeH}) includes $\ddot{H}$, we can apply their results in the slow-roll approximation under the conditions (\ref{slowroll}) after the identification of the parameters as $\a l^2= 2/m^2$, where $\a$ is the dimensionless coupling constant introduced in Ref.~\cite{Cano:2020oaa} and $l=(\a')^{1/2}$ is the fundamental length in superstring theory.
Equation~\eqref{EQU00} can be put to the form
\be\label{ser}
R \left( \fracmm{R}{12}-{H}^{2} \right) -H{\dot{R}}=3\,{m}^{2} \left( H^2 -\fracmm{3\b}{m^4}H^6 - \fracmm{3\b}{m^6}H^8\right)\equiv 3m^2 F(H^2)~,
\ee
so that the effective $F(H^2)$ function of Ref.~\cite{Cano:2020oaa} in our case is given by
\be \lb{ourF}
F(H^2) = H^2 - \fracmm{3\b}{m^4} \left(H^2\right)^3 - \fracmm{3\b}{m^6} \left(H^2\right)^4~.
\ee
Since $\b >0$, it corresponds to the $F_5$-scenario in the classification of Ref.~\cite{Cano:2020oaa}.

Accordingly, we have
\be \lb{Ffder}
F'(H^2) = 1 - 9\b \left(\fracmm{H}{m}\right)^4 - 12\b \left(\fracmm{H}{m}\right)^6
\ee
and
\be \lb{Fsder}
F''(H^2) ={} -18 \b \fracmm{H^2}{m^4} - 36\b \fracmm{H^4}{m^6}~~,
\ee
where the primes here denote the differentiations with respect to $H^2$.

As was demonstrated in Ref.~\cite{Bueno:2016ypa}, the effective Newton constant in the  higher-derivative gravities
must obey the condition (in the notation adapted to the $F$-function of Ref.~\cite{Cano:2020oaa})
\be \lb{effN}
G_{\rm{eff.}} = \fracmm{1}{8\pi M_{\rm{Pl}}^2\left[ F'(H^2) + 4(H^2/m^2)\right]}>0
\ee
in order to avoid graviton ghosts. Given the $F$-function (\ref{ourF}), we find the restriction
\be \lb{firstconF}
1+ 4h^2 - 9\b h^4 - 12\b h^6 >0~.
\ee
When using the upper bound on $h$ in Eq.~(\ref{maxh}), Eq.~(\ref{firstconF}) implies
\be
\lb{maxb1}
\b <  6.941\cdot 10^{-4}\,.
\ee

A stronger  condition was also proposed in Ref.~\cite{Cano:2020oaa} by demanding the absence of negative energy fluxes,
which is required by unitarity and causality constraints. This condition reads~\cite{Cano:2020oaa}
\be \lb{nonegf}
-4 \leqslant \fracmm{210 H^2 F''(H^2)}{F'(H^2) + 4(H^2/m^2)}\leqslant  4~~.
\ee
In our case (\ref{ourF}), it leads to the condition
\be\lb{sconFmod}
\beta \leqslant \fracmm{1+4h^2}{6h^4 \left( 317 h^2+159 \right) }
\ee
that is  satisfied when
\be \lb{maxb2}    \b \leqslant  4.428\cdot 10^{-6}~~,
\ee
where we have used $h\leqslant 4.655$ from Eq.~(\ref{maxh}).

Yet another upper bound on $\beta$ can be obtained from CMB measurements. According to the CMB observation data  \cite{Planck:2018jri,BICEP:2021xfz,Tristram:2021tvh}, we have
\be  \lb{smbt}
n_s = 0.9649\pm 0.0042~,\quad  r < 0.036~~,
\ee
for the tilt $n_s$ of scalar (curvature) perturbations and the tensor-to-scalar ratio $r$.

The results of Ref.~\cite{Cano:2020oaa} for the observable CMB tilts, specified to our case, are given by
\be  \lb{nsF}
n_s=1-\fracmm{2}{N}-\fracmm{8\b N}{9}-\fracmm{\b N^2}{2}
\ee
and
\be
r=\fracmm{12}{N^2}-\fracmm{16}{3}\b -2\b N~,
\ee
for scalar and tensor perturbations, respectively. The quantum correction to the Starobinsky inflation model in Eq.~(\ref{sbrm}) decreases both tilts $n_s$ and $r$ because $\beta>0$. 

\begin{figure}[h!tbp]
\includegraphics[width= 0.9\textwidth]{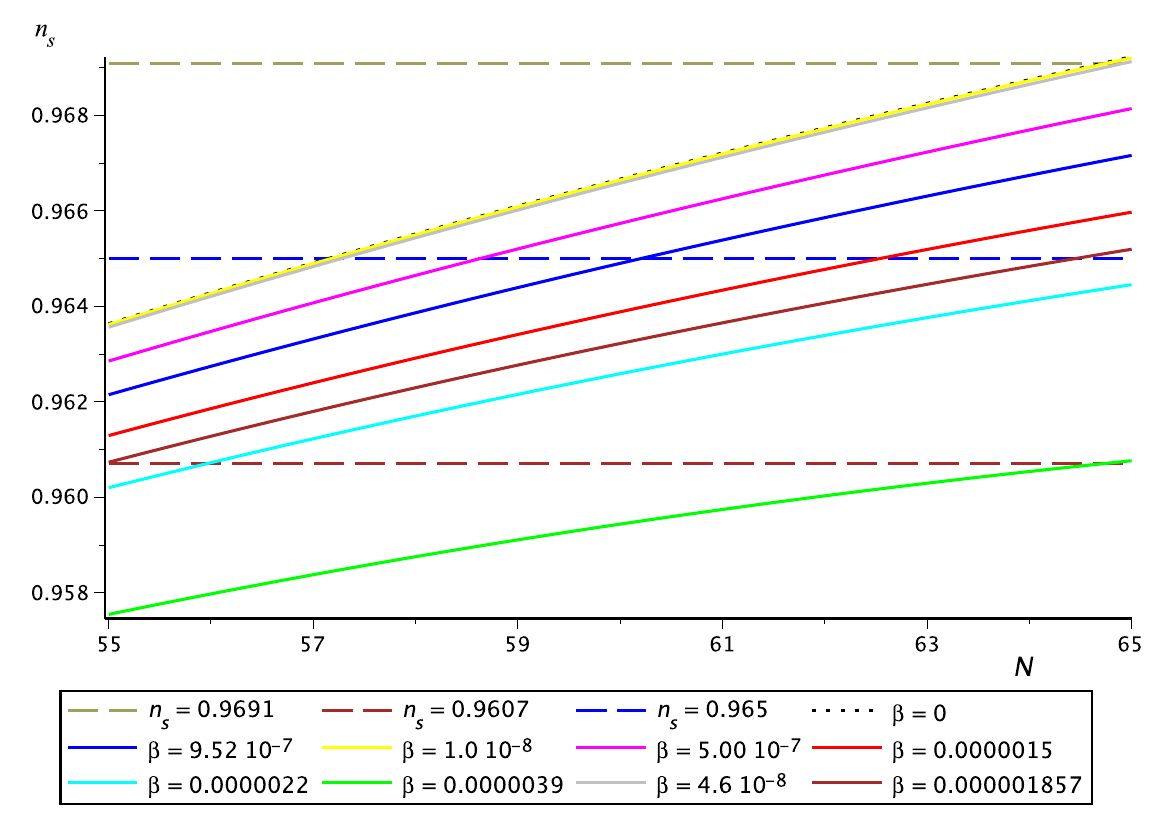}
\caption{The spectral index $n_{s}$ for $0\leqslant \beta \leqslant 3.9\cdot10^{-6}$ with the e-foldings $55\leqslant N\leqslant 65$. The dotted lines are the boundaries for  the observed value of $n_s$ set by the CMB data.}
\label{Picn}
\end{figure}

\begin{figure}[h!tbp]
\includegraphics[width= 0.9\textwidth]{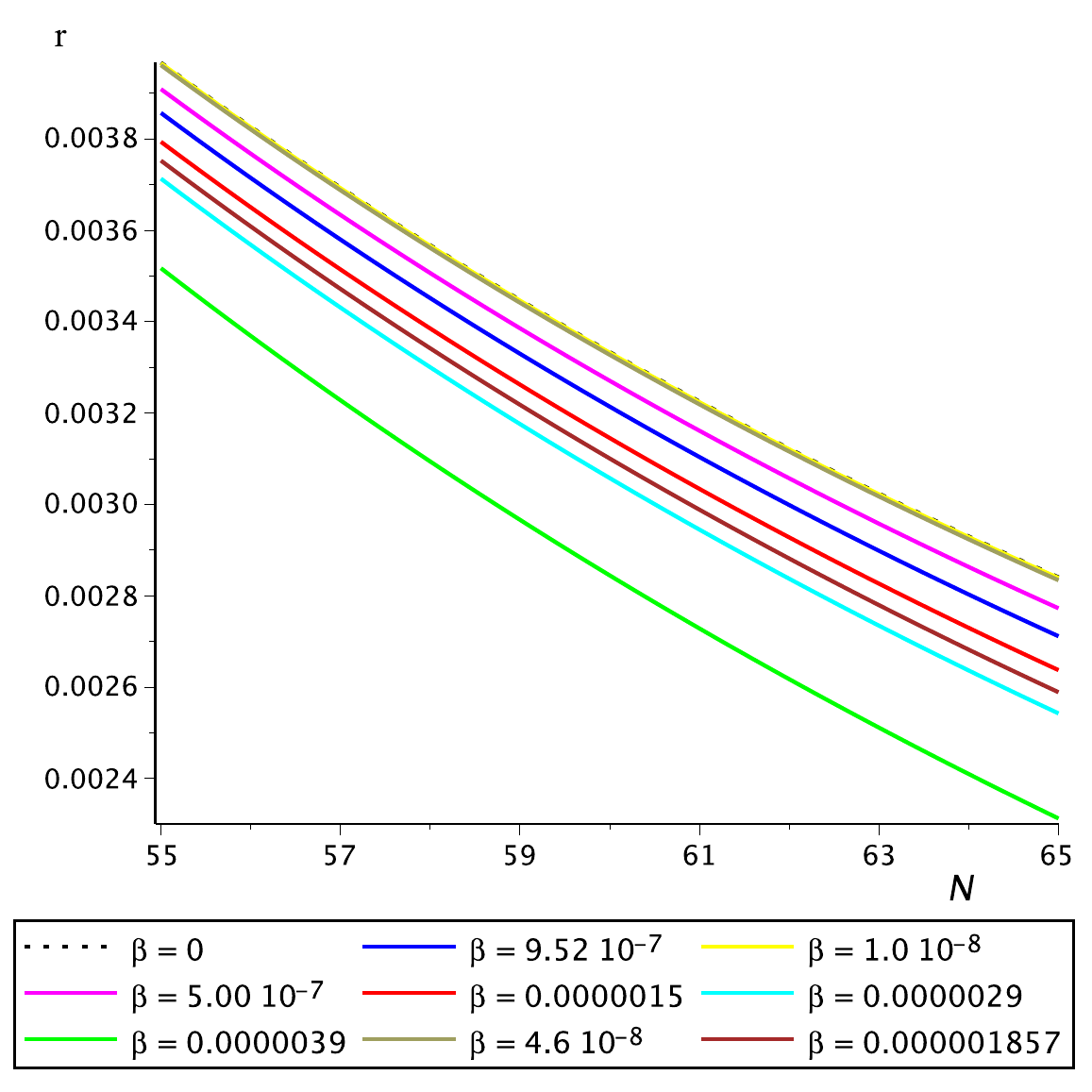}
\caption{The  tensor-to-scalar ratio $r$ for $0\leqslant \beta \leqslant 3.9\cdot10^{-6}$ and e-folds $55\leqslant N\leqslant 65$.}
\label{Picr}
\end{figure}

The inflationary parameters as the functions of $N$ are presented in Figs.~\ref{Picn} and \ref{Picr} for some values of $\beta$.
For example,\\ (i) to get $n_s=0.9691$ with $N=65$, we need $\beta=4.608\cdot 10^{-8}$,\\
(ii) to get $n_s=0.9607$ with $N=55$, we need $\beta=1.857\cdot10^{-6}$, and \\ (iii)
to get $n_s=0.9607$ with $N=65$,  we need $\beta=3.9\cdot10^{-6}$.

Therefore, in order to be consistent with the observed value of the spectral index $n_s$ for all $55<N<65$, we should demand
\be
\lb{betaUB}
\b \leqslant 3.9\cdot10^{-6}~~.
\ee
The tensor-to-scalar-ratio $r$ is under the upper bound (\ref{smbt}) for these values of $\beta$, see Fig.~\ref{Picr}.

In order to calculate the $\beta$-correction to the observable (CMB) amplitude $A_s$ of scalar perturbations, we take the slow-roll parameter
$\epsilon$ in terms of the function $h^2(N)$,
\bea
\epsilon=\fracmm{1}{2}\fracmm{d\ln h^2}{dN}~,
\eea
and use our solution (\ref{hbeta}) for $h(N)$. We find
\bea
 \epsilon&\approx&\fracmm{1}{2N}+\fracmm{\beta\,N}{3}\left(\fracmm{N}{8}+\fracmm{1}{3} \right)
\eea
and
\be
A_s=\fracmm{(1+\zeta/9)h^2}{16\,\pi^2\,\epsilon }\fracmm{m^2}{M^2_{Pl}}~~,
\ee
where we have introduced the new parameter $\zeta$ as
\be\zeta={}-{\fracmm {9\,(4\,\epsilon+n_s-1)}{8\,\epsilon}}\approx\fracmm{\beta\,{N}^{2} \left( 3\,N+4 \right) }{4}~~.
\ee

Therefore, we get
\be
\lb{amplitude}
 A_s=\left({\fracmm{{N}^{2}}{24\,{\pi }^{2}}}+\fracmm{{N}^{5}\beta}{864\,\pi^2}\right)\fracmm{m^2}{M_{\rm Pl}^2}~~,
\ee
where the first term is standard and has the value $\bar{A}_s= 2.1\cdot10^{-9}$ for the best fit $N=55$.

Substituting  $m=1.3(\fracmm{55}{N}) 10^{-5}M_{\rm Pl}$ into Eq.~(\ref{amplitude}) gives the $\beta$-correction and
\be
A_s\approx 2.1\cdot10^{-9}+5.5\cdot 10^{-11}N^3\beta~~.
\ee
For instance, when $N=65$ and $\beta=10^{-6}$, we get  the $\beta$-correction of the order ${\cal O}(10^{-3}) \bar{A}_s$.

\section{Conclusion}

The Starobinsky model of inflation~\cite{Starobinsky:1980te} is in very good agreement with the current observational data of the cosmic microwave background radiation~\cite{Planck:2018jri,BICEP:2021xfz}. However,  the Starobinsky inflation solution is unstable against adding {\it generic} terms of the higher-order in the spacetime curvature. In this paper, we studied physical applications of the Bel-Robinson tensor $T^{\mu\nu\lambda\rho}$ squared term, proposed as the leading quantum correction inspired by superstring theory, to the inflationary stage of the early universe evolution.~\footnote{Physical applications of the SBR gravity to Schwarzschild-type black holes, as regards their Hawking temperature, entropy, pressure and lifetime in the first order with respect to $\b$, are given in Ref.~\cite{CamposDelgado:2022sgc}.}  The proposed gravitational EFT  action includes squares of two topological densities $E_4={\cal G}$ and $P_4$. The $P_4^2$ term does not contribute to the evolution equations in a spatially flat FLRW universe, so that the action reduces to the particular case of the $F(R, \cal{G})$ modified gravity on the FLRW background. Remarkably, the positivity of the new parameter $\b$ is a necessary condition for cosmological viability of the Einstein-Gauss-Bonnet models~\cite{DeFelice:2008wz}, while it is  also a consequence of the origin of the Bel-Robinson tensor squared term from superstrings~\cite{Ketov:2022lhx}.

The string-inspired $\beta$-correction in the action (\ref{sbr})  leads to the modification of the Starobinsky equation (\ref{eomH})
for the Hubble function, which is given by Eq.~(\ref{odeH}). Unlike Eq.~(\ref{eomH}) that has the quasi-de Sitter solution,  Eq.~(\ref{odeH}) also has the exact de Sitter solution given by our Eq.~(\ref{deS}). Since $3\b\leqslant {\cal O}(10^{-5})$ according to our results in Sec.~5, the value of $H_{\rm dS}$ is above ${\cal O}(10^{-4})M_{\rm Pl}$, where we have used Eq.~(\ref{mass}). Therefore, it implies the existence of a very short de Sitter phase before inflation. The de Sitter solution is not an attractor and is unstable against perturbations of $H_{\rm dS}$. Whether the de Sitter phase really happened before inflation or not depends upon non-perturbative considerations in superstring/M-theory, which is beyond the scope of this paper.

Since we extended the Starobinsky inflation model by the new parameter $\b$, the predictions for CMB observables are modified. We obtained the leading corrections in the first order with respect to $\beta$ and the physical bounds of the parameter $\beta$ in Sec.~5.

The next generation of CMB experiments e.g., the satellite missions LiteBIRD \cite{LiteBIRD:2022cnt} and CORE \cite{CORE:2016ymi}, as well as the ground-based experiments POLARBEAR \cite{POLARBEAR:2019kzz},
BICEP/Keck \cite{BICEP:2021xfz} and Simons Observatory \cite{SimonsObservatory:2018koc}, will measure the values of the cosmological tilts and the CMB amplitude with higher precision, which may also probe quantum corrections to the Starobinsky inflation.

\section*{Acknowledgements}

The authors are grateful to Alexei A. Starobinsky for useful discussions and correspondence, and thank Ruben C. Delgado and Shunsuke Toyama for their help in verifying some equations with Mathematica.

S.V.K. was supported by Tokyo Metropolitan University, the Japanese Society for Promotion of Science under the grant No.~22K03624,
the Tomsk Polytechnic University development program Priority-2030-NIP/EB-004-0000-2022, and the World Premier International Research Center Initiative (WPI Initiative), MEXT, Japan.  E.O.P. and S.Yu.V. were partially supported by the Russian Foundation for Basic Research grant No.~20-02-00411.

\begin{appendix}

\section{Riemann geometry notation and useful formulae}
\label{Notations}
We use the natural units $c=\hbar=1$ in a four-dimensional curved spacetime with local coordinates $x^{\m}$ and the signature $(-,+,+,+)$. The basic equations of Riemann geometry with a metric $g_{\m\n}(x)$ in our notation are (we use lower case Greek letters for $D=4$ curved spacetime vector indices $\m,\n,\ldots=0,1,2,3$)
\begin{eqnarray}
    & & ds^2 =g_{\m\n} dx^{\m}dx^{\n}~,\nonumber\\
   & & \Gamma^{\alpha}_{\mu\nu}=\fracmm{1}{2}g^{\alpha\sigma}(\partial_\nu g_{\mu\sigma}+\partial_\mu g_{\nu\sigma}-\partial_\sigma g_{\mu\nu}),
   \nonumber\\
    & & [\nabla_\alpha,\nabla_\beta]A^\mu = R^\mu_{\,\,\,\sigma\alpha\beta}A^\sigma~,\nonumber\\
    & & R^{\rho}_{\,\,\,\sigma\mu\nu}=\partial_\mu\Gamma^{\rho}_{\sigma\nu}-\partial_\nu\Gamma^{\rho}_{\sigma\mu}+ \Gamma^{\rho}_{\mu\lambda}\Gamma^{\lambda}_{\nu\sigma}-\Gamma^{\rho}_{\nu\lambda}\Gamma^{\lambda}_{\mu\sigma}~~,\nonumber\\
   & & R_{\mu\nu}=R^{\alpha}_{\,\,\,\mu\alpha\nu}=\,\partial_\sigma\Gamma^{\sigma}_{\mu\nu}-\partial_\nu\Gamma^{\sigma}_{\mu\sigma}+
\Gamma^{\sigma}_{\mu\nu}\Gamma^{\rho}_{\sigma\rho}-\Gamma^{\rho}_{\sigma\nu}\Gamma^{\sigma}_{\mu\rho}~,\nonumber\\
   & & R=g^{\sigma\tau}R_{\sigma\tau}\,=\,R^{\sigma}_{\sigma}\,=\,g^{\tau\xi}\Gamma^{\sigma}_{\tau\xi,\sigma}-g^{\tau\xi}
\Gamma^{\sigma}_{\tau\sigma,\xi}+g^{\tau\xi}\Gamma^{\sigma}_{\tau\xi}\Gamma^{\rho}_{\sigma\rho}-g^{\tau\xi}\Gamma^{\rho}_
{\tau\sigma}\Gamma^{\sigma}_{\xi\rho}~,\nonumber\\
 & & \sqrt{-g}\nabla_\mu A^{\mu}=\partial_\mu(\sqrt{-g}A^\mu)~,\nonumber\\
   & & [\nabla_\alpha,\nabla_\beta]T^\mu_\lambda = R^\mu_{\,\,\,\sigma\alpha\beta}T^\sigma_\lambda-R^\sigma_{\,\,\,\lambda\alpha\beta}T^\mu_\sigma~,\nonumber\\
   & &\Box=\nabla^\mu\nabla_\mu,
    \end{eqnarray}
where $\pa_{\m}=\pa/\pa {x^{\m}}$,  $A^\mu$ is an arbitrary vector field and $T^\mu_\lambda$ is an arbitrary rank-two tensor field. As regards a
generic tensor field, we have
\begin{equation}
\begin{split}
\nabla_{\r}T^{\m_1\m_2\ldots \m_p}_{\quad\quad\quad \n_1\n_2\ldots \n_q} & =  \partial_{\r}T^{\m_1\m_2\ldots\m_p}_{\quad\quad \quad \n_1\n_2\ldots\n_q}
+ \sum_{k=1}^{p}T^{\m_1\ldots \l_k\ldots \m_p}_{\quad\quad\quad ~\n_1\n_2\ldots\n_q}\Gamma^{\m_k}_{\r\l_k} \\
 & {}-\sum_{m=1}^{q}T^{\m_1\m_2\ldots\m_p}_{\quad\quad\quad ~\n_1\ldots \l_m\ldots \n_q}\Gamma^{\l_m}_{\r\n_m} ~.
\end{split}
\end{equation}

The algebraic Bianchi identities are
\be  \label{aBI}
R_{\m\n\l\r} =- R_{\n\m\l\r}= R_{\m\n\r\l}= R_{\l\r\m\n}~,\quad R_{\m\n\l\r} +   R_{\r\m\n\l}+ R_{\l\r\m\n}  =0~.
\ee
The differential Bianchi identities are
 \begin{eqnarray} \label{dBI}
   & & \nabla_\mu R^\mu_\nu=\fracmm{1}{2}\nabla_\nu R=\fracmm{1}{2}\partial_\nu R~,\nonumber\\
   & & \nabla_\mu R^{\mu}_{\,\,\,\,\,\nu\lambda\rho}=\nabla_\lambda R_{\nu\rho}-\nabla_\rho R_{\nu\lambda}~,\nonumber\\
   & & \nabla_\sigma R_{\mu\nu\alpha\beta}+\nabla_\nu R_{\sigma\mu\alpha\beta}+\nabla_\mu R_{\nu\sigma\alpha\beta}=0~.
\end{eqnarray}

\section{Euler density and Gauss-Bonnet-term}
\label{EGB}
By using the identity
 \be\label{epsilon}\epsilon_{\a\b\g\d}\epsilon^{\m\n\l\r}=\left|
                                                    \begin{array}{cccc}
                                                      \delta^{\m}_{\a} & \delta^{\m}_{\b} & \delta^{\m}_{\g} & \delta^{\m}_{\d} \\
                                                      \delta^{\n}_{\a} & \delta^{\n}_{\b} & \delta^{\n}_{\g} & \delta^{\n}_{\d} \\
                                                      \delta^{\l}_{\a} & \delta^{\l}_{\b} & \delta^{\lambda}_{\g}  & \delta^{\l}_{\d} \\
                                                      \delta^{\r}_{\a} & \delta^{\r}_{\b} & \delta^{\r}_{\g} & \delta^{\r}_{\d} \\
                                                    \end{array}
                                                   \right|\end{equation}
and the algebraic Bianchi identities (\ref{aBI}), it is straightforward to verify that
\be
\label{E4G}
E_4={}^*R_{\m\n\l\r}{}^*R^{\m\n\l\r}=R^2-4R_{\alpha\beta}R^{\alpha\beta}+R_{\alpha\beta\gamma\delta}R^{\alpha\beta\gamma\delta}={\cal G}~.
\end{equation}

\section{Field variations of geometrical quantities}
\label{Equa}
It is straightforward to get  the basic equations for metric variations,
\begin{eqnarray}\label{variationalcalculus}
\begin{array}{ll}
 \delta g^{\mu\nu}={}-g^{\mu\rho}g^{\nu\eta}\delta g_{\rho\eta}~,\\
 g_{\rho\lambda}\delta g^{\kappa\lambda}={}-g^{\kappa\lambda}\delta g_{\rho\lambda}~,\\
  \delta\sqrt{-g}={}-\fracmm{\sqrt{-g}}{2}g_{\mu\lambda}\delta g^{\mu\lambda} \label{deltag}~,\\
\delta \Gamma^\mu_{\lambda\beta} =\fracmm{1}{2} g^{\mu\rho}(\nabla_\lambda\delta g_{\beta\rho}+\nabla_\beta\delta g_{\lambda\rho}-\nabla_\rho\delta g_{\lambda\beta})~,  \\
\delta R^\mu_{\lambda\alpha\beta}=\nabla_\alpha(\delta \Gamma^\mu_{\lambda\beta})-\nabla_\beta(\delta\Gamma^\mu_{\lambda\alpha})~, \\
\delta R_{\sigma\nu}=\fracmm{1}{2}\left(-g_{\nu\lambda}\nabla_\mu\nabla_\sigma-g_{\sigma\lambda}\nabla_\mu\nabla_\nu+g_{\sigma\mu}g_{\nu\lambda}\Box+g_{\mu\lambda}\nabla_\nu\nabla_\sigma\right)\delta g^{\mu\lambda}~,\\
g^{\mu\nu}\delta R_{\mu\nu}={}-\nabla_\mu \nabla_\nu\delta g^{\mu\nu}+\Box (g_{\mu\nu}\delta g^{\mu\nu})~,\\
\delta (R_{\alpha\beta}R^{\alpha\beta})\,=\,2R_\mu^{\,\,\,\,\alpha}R_{\alpha\nu}\delta
g^{\mu\nu}+2R^{\mu\nu}\delta R_{\mu\nu}~,\\
\delta R^{\mu\nu}_{\quad\alpha\beta}= R^{\mu}_{\,\,\lambda\alpha\beta}\,\delta g^{\lambda \nu}+ g^{\lambda \nu}\delta R^{\mu}_{\,\,\lambda\alpha\beta}~,\\
\delta\,(R_{\alpha\beta\gamma\delta}R^{\alpha\beta\gamma\delta})\,=\delta(R_{\alpha\beta}^{\quad\gamma\delta}
R^{\alpha\beta}_{\quad\gamma\delta})=2\,R_{\alpha\beta}^{\quad\gamma\delta}\delta R^{\alpha\beta}_{\quad\gamma\delta}~.
\\
\end{array}
\end{eqnarray}

A variation of the curvature tensor reads
 \be
 {\delta R^{\g\d}_{\quad\alpha\beta}}={\delta g^{\lambda \d}}R^{\g}_{\,\,\lambda\alpha\beta}+g^{\lambda \d}{\delta R^{\g}_{\,\,\lambda\alpha\beta}}~,
 \end{equation}
 where
 \nbe\delta R^{\d}_{\,\,\lambda\alpha\beta}=\nabla_\alpha\left(\delta \Gamma^{\d}_{\lambda\beta}\right)-\nabla_\beta \left(\delta
 \Gamma^{\d}_{\lambda\alpha}\right)
 \nee
\nbe
{}=\fracmm{g^{\d\rho}}{2}(\nabla_\alpha\nabla_\beta-\nabla_\beta\nabla_\alpha)\delta g_{\lambda\rho}+\fracmm{g^{\d\rho}}{2}\left(\nabla_\alpha\nabla_\lambda\delta g_{\rho \beta}-\nabla_\alpha\nabla_\rho\delta g_{\lambda \beta}\right)-\fracmm{g^{\d\sigma}}{2}\left(\nabla_\beta\nabla_\lambda\delta g_{\sigma\alpha}-\nabla_\beta\nabla_\sigma\delta g_{\lambda\alpha}\right)~.
\nee
It follows that
 \be
 \delta R^{\g\d}_{\quad\alpha\beta}=\fracmm{1}{2} \left( R^{\g}_{\lambda\alpha\beta}\delta g^{\lambda \d}-R^{\d}_{\lambda\alpha\beta}\delta g^{\lambda \g}\right)+\fracmm{1}{2}\left(g_{\beta\lambda}\nabla_\alpha-g_{\alpha\lambda}\nabla_\beta\right)\left(\nabla^{\g}\delta g^{\lambda \d}-\nabla^{\d}\delta g^{\lambda \g}\right)~.
 \end{equation}

Accordingly, a variation of the GB term is given by
\begin{equation*}
    \delta {\cal G}=2R\delta R-8R^\alpha_\mu \delta R^\mu_\alpha+2R_{\mu\nu}^{\quad\alpha\beta}\delta  R^{\mu\nu}_{\quad\alpha\beta}=2R\delta R-8R^{\alpha\beta}\delta R_{\alpha\beta}-8R^\sigma_{\nu}R_{\sigma\beta}\delta g^{\nu\beta}+2R_{\alpha\beta}^{\quad \mu\nu}\delta R^{\alpha\beta}_{\quad\mu\nu}~,
\end{equation*}
where we find
\begin{equation*}
\delta R=R_{\mu\nu}\delta g^{\mu\nu}+g^{\mu\nu}\delta R_{\mu\nu}=\left[R_{\mu\nu}-\nabla_\mu \nabla_\nu+\Box g_{\mu\nu}\right]\delta g^{\mu\nu},
\end{equation*}
\begin{equation*}
    \delta R^\mu_\alpha=\fracmm{1}{2}\left[R^\mu_{\lambda\alpha\beta}\delta g^{\lambda\beta}+R_{\lambda\alpha}\delta g^{\lambda\mu}+g_{\beta\lambda}\nabla_\alpha\nabla^\mu\delta g^{\lambda\beta}-g_{\alpha\lambda}\nabla_\beta\nabla^\mu\delta g^{\lambda\beta}-\nabla_\alpha\nabla_\lambda\delta g^{\lambda\mu}+g_{\alpha\lambda}\Box\delta g^{\lambda\mu}\right]
\end{equation*}
\begin{equation}\label{deltaRmunu2}
    \delta R^{\mu\nu}_{\quad\alpha\beta}=\fracmm{1}{2}\left[R^\mu_{\,\,\,\lambda\alpha\beta}\delta g^{\lambda\nu}-R^\nu_{\,\,\,\lambda\alpha\beta}\delta g^{\lambda\mu}+
    (g_{\beta\lambda}\nabla_\alpha-g_{\alpha\lambda}\nabla_\beta)(\nabla^\mu\delta g^{\lambda\nu}-\nabla^\nu\delta g^{\lambda\mu})\right].
\end{equation}
Taking these contributions together, we get
\begin{equation}\label{deltaG}
\begin{split}
    \delta {\cal G}=&\left[2RR_{\rho\nu}-8R_{\rho\alpha}R^\alpha_\nu+2R_{\rho\mu\alpha\beta}R_\nu^{\,\,\,\mu\alpha\beta}+2(Rg_{\rho\nu}-2R_{\rho\nu})\Box-R(\nabla_\nu\nabla_\rho+\nabla_\rho\nabla_\nu)\right.\\
        +&\left.4(R^\alpha_\nu\nabla_\rho\nabla_\alpha+R^\alpha_\rho\nabla_\nu\nabla_\alpha)-4g_{\rho\nu}R_{\alpha\beta}\nabla^\alpha\nabla^\beta-2R_{\alpha\rho\nu\mu}(\nabla^\mu\nabla^\alpha+\nabla^\alpha\nabla^\mu)\right]\delta g^{\rho\nu}~.
\end{split}
\end{equation}

After substituting this variation into the BSR action (\ref{Gaction}) and integrating by parts we find
\begin{equation}
\begin{split}
    &\int d^4x\chi\delta(\sqrt{-g}{\cal{G}})=2\int d^4x\sqrt{-g}\left[(Rg_{\rho\nu}-R_{\rho\nu})\Box\chi-R\nabla_\rho\nabla_\nu\chi\right.\\
    &\left. { } + 2\left(R^\alpha_\nu\nabla_\alpha\nabla_\rho\chi+R^\alpha_\rho\nabla_\alpha\nabla_\nu\chi\right)-2\left(g_{\rho\nu}R_{\alpha\beta}+R_{\alpha\rho\nu\beta}\right)\nabla^\beta\nabla^\alpha\chi\right]\delta g^{\rho\nu}~.
\end{split}
\end{equation}

\section{Painlev\'e tests}
\label{Painleve}

The Painlev\'e analysis~\cite{Miritzis:2000js,Gorielydoi:10.1142/3846,Conte} is used in a  search for integrable cosmological models in modified gravity~\cite{Miritzis:2000js,Paliathanasis:2016tch,Leon:2022dwd}.
It was demonstrated in Ref.~\cite{Paliathanasis:2017apr} that the Starobinsky model passes the so-called weak Painlev\'{e} test~\cite{Miritzis:2000js}.

A generic solution to Eq.~(\ref{eomH}) can be written in the form of Puiseux series as follows:
\begin{equation}
\label{2exp}
\begin{split}
 H(t) & =  \fracmm{1}{2(t-t_0)}+C\sqrt{t-t_0}-\fracmm{m^2}{6}(t-t_0)+\fracmm{C^2}{8}(t-t_0)^2-\fracmm{m^2C}{12}(t-t_0)^{5/2} \\
 &{}+  \fracmm{2m^4}{225}(t-t_0)^{3} -\fracmm{23}{88}C^3(t-t_0)^{7/2}+\fracmm{5m^2C^2}{42}(t-t_0)^4-\fracmm{1627m^4C}{93600}(t-t_0)^{9/2}\\
 &{}+  \left[\fracmm{4}{4725}m^6 +\fracmm{457C^4}{2464}\right](t-t_0)^5-\fracmm{655m^2C^3}{7392}(t-t_0)^{11/2}+
 \fracmm{441797m^4C^2}{28828800}(t-t_0)^6  \\
 &{}-\fracmm{\left(333311m^6+21012615C^4\right)C}{294053760}(t-t_0)^{13/2}+\left[\fracmm{46}{1535625}m^8+\fracmm{725m^2C^4}{24024}\right](t-t_0)^7+\dots\!,
\end{split}
\end{equation}
where $t_0$ and $C$ are two constants of integration.

When $(t-t_0) >0$, a real solution requires a real $C$, whereas when
$(t-t_0) <0$, a real solution requires a pure imaginary $C$. The expansion (\ref{2exp}) is valid for $m\abs{t-t_0}\ll 1$.

The Painlev\'e test is a procedure for investigating the leading-order behavior of the general solution in the neighborhood of its singularity. It leads to resonances related to the appearance of integration constants. The Painlev\'e test in the Kowalevskaya form consists of three steps (the methodology of the Painlev\'e analysis is described in Refs.~\cite{Miritzis:2000js,Gorielydoi:10.1142/3846,Conte}).
First, we use a simple {\it ansatz} for a singular solution to Eq.~(\ref{odeH}) in the form
\be\lb{ans1}
H=\fracmm{c_{-p}}{t^p}
\ee
with constants $c_{-p}$ and $p>0$ to be defined. Substituting Eq.~(\ref{ans1}) into Eq.~(\ref{odeH}) with a nonzero parameter $\beta$ and  keeping
only the most singular terms, we get the equation
\begin{equation}
\label{EQU00mostsingular}
2{H}^{5}{\ddot{H}}+3{H}^{4}{\dot{H}}^{2}+6{H}^{6}{\dot{H}}-{H}^{8}=0,
\end{equation}
and find that Eq.~\eqref{EQU00mostsingular} is satisfied when $p=1$ and
either $c_{-1}=1$ or $c_{-1}=-7$.

Second, by using the results obtained, we extend our {\it ansatz} for a solution to  Eq.~\eqref{EQU00mostsingular} to the
form
\begin{equation} \lb{ans2}
           H=\fracmm{c_{-1}}{t}+Ct^{n-1}
\end{equation}
with new unknown constants $C$ and $n$. Substituting the ansatz (\ref{ans2}) into Eq.~\eqref{EQU00mostsingular}, we find
that the terms linear in $C$ vanish, while $n$ is a solution to the quadratic equation, whose two roots are given by $n_1=-1$ and
$n_2=4$ at $c_{-1}=1$, and by $n_1=-1$ and $n_2=28$ at $c_{-1}=-7$. It means that the resonances can arise in the terms proportional to $t^3$ for $c_{-1}=1$ and in those proportional to $t^{27}$ for $c_{-1}=-7$.

Third, we check the existence of resonances. In the case of $c_{-1}=1$ and $n_2=4$, we substitute  the ansatz
\begin{equation}
H(t)=\fracmm{1}{t}+c_{0}+c_{1}t+c_{2}t^2+c_{3}t^3
\end{equation}
into Eq.~(\ref{odeH}), we find that there is no solution in the form of such Laurent series.

When $c_{-1}=-7$ and $n_2=28$, the corresponding {\it ansatz} for a solution to Eq.~(\ref{odeH}) reads
\begin{equation} \lb{Ansatz}
H(t)={}-\fracmm{7}{t}+\sum\limits_{k=0}^{27}c_{k}t^k
\end{equation}
with unknown coefficients $c_k$. Substituting Eq.~(\ref{Ansatz}) into Eq.~(\ref{odeH}), we find a solution in the form of the Laurent series with the leading terms given by
\begin{equation}
H(t)={}-\fracmm{7}{t}+\fracmm{m^4}{5488\beta}t^3+\fracmm{m^6}{316932\beta}t^5+\fracmm{919m^8}{45538633728\beta^2}t^7+\ldots~.
\end{equation}

The vanishing sum of the terms proportional to $t^{27}$ in Eq.~(\ref{odeH}) leads to a
quadratic equation on $\beta$,
\begin{equation}
8076977541353570304\beta^2-29100733749549586432\beta+2230185058312760175=0,
\end{equation}
that has two real solutions,
\begin{equation}
\beta_{+}= {}\fracmm{5918093565659}{3285161757696}-\fracmm{121}{1642580878848}\sqrt{547160457121493937679}\approx 0.078340116
\end{equation}
and
\begin{equation}
 \beta_{-} ={} \fracmm{5918093565659}{3285161757696}+\fracmm{121}{1642580878848}\sqrt{547160457121493937679}\approx 3.524583576.
\end{equation}
We conclude that Eq.~(\ref{odeH}) passes the Painlev\'e test only for two values of $\beta$, namely, $\beta_{+}$ and $\beta_{-}$. These values of $\beta$ are positive, so, the corresponding Einstein--Gauss--Bonnet gravity models are cosmologically viable~\cite{DeFelice:2008wz}. However, these values of $\beta$ are too large for viable inflation.

\end{appendix}

\bibliography{Bibliography}{}
\bibliographystyle{JHEP}
\end{document}